\documentclass[twocolumn,superscriptaddress,amsmath,amssymb,prx,floatfix]{revtex4-2}

\usepackage{xr}
\usepackage{hyperref}
\hypersetup{
    colorlinks=true,
    linkcolor=blue,
    filecolor=magenta,      
    urlcolor=cyan,
    pdfpagemode=FullScreen,
    }

\usepackage{graphicx}
\usepackage{dcolumn}
\usepackage{bm}

\usepackage[utf8]{inputenc}
\usepackage[T1]{fontenc}
\usepackage{mathptmx}
\usepackage[dvipsnames]{xcolor}
\usepackage{geometry}
\usepackage[explicit]{titlesec}

\usepackage{xcolor}
\usepackage[draft]{changes}
\definechangesauthor[color=red, name={Paulo Santos}]{PVS}
\definechangesauthor[color=cyan, name={Alexander Kuznetsov}]{AK}

\usepackage{verbatim}
\setauthormarkup{}

\RequirePackage[normalem]{ulem} 
\RequirePackage{color}\definecolor{RED}{rgb}{1,0,0}\definecolor{BLUE}{rgb}{0,0,1} 
\providecommand{\DIFaddbegin}{} 
\providecommand{\DIFaddend}{} 
\providecommand{\DIFdelbegin}{} 
\providecommand{\DIFdelend}{} 
\providecommand{\DIFaddbeginFL}{} 
\providecommand{\DIFaddendFL}{} 
\providecommand{\DIFdelbeginFL}{} 
\providecommand{\DIFdelendFL}{} 
\newcommand{\DIFscaledelfig}{0.5}
\RequirePackage{settobox} 
\RequirePackage{letltxmacro} 
\newsavebox{\DIFdelgraphicsbox} 
\newlength{\DIFdelgraphicswidth} 
\newlength{\DIFdelgraphicsheight} 
\LetLtxMacro{\DIFOincludegraphics}{\includegraphics} 
\newcommand{\DIFaddincludegraphics}[2][]{{\color{blue}\fbox{\DIFOincludegraphics[#1]{#2}}}} 
\newcommand{\DIFdelincludegraphics}[2][]{
\sbox{\DIFdelgraphicsbox}{\DIFOincludegraphics[#1]{#2}}
\settoboxwidth{\DIFdelgraphicswidth}{\DIFdelgraphicsbox} 
\settoboxtotalheight{\DIFdelgraphicsheight}{\DIFdelgraphicsbox} 
\scalebox{\DIFscaledelfig}{
\parbox[b]{\DIFdelgraphicswidth}{\usebox{\DIFdelgraphicsbox}\\[-\baselineskip] \rule{\DIFdelgraphicswidth}{0em}}\llap{\resizebox{\DIFdelgraphicswidth}{\DIFdelgraphicsheight}{
\setlength{\unitlength}{\DIFdelgraphicswidth}
\begin{picture}(1,1)
\thicklines\linethickness{2pt} 
{\color[rgb]{1,0,0}\put(0,0){\framebox(1,1){}}}
{\color[rgb]{1,0,0}\put(0,0){\line( 1,1){1}}}
{\color[rgb]{1,0,0}\put(0,1){\line(1,-1){1}}}
\end{picture}
}\hspace*{3pt}}} 
} 
\LetLtxMacro{\DIFOaddbegin}{\DIFaddbegin} 
\LetLtxMacro{\DIFOaddend}{\DIFaddend} 
\LetLtxMacro{\DIFOdelbegin}{\DIFdelbegin} 
\LetLtxMacro{\DIFOdelend}{\DIFdelend} 
\DeclareRobustCommand{\DIFaddbegin}{\DIFOaddbegin \let\includegraphics\DIFaddincludegraphics} 
\DeclareRobustCommand{\DIFaddend}{\DIFOaddend \let\includegraphics\DIFOincludegraphics} 
\DeclareRobustCommand{\DIFdelbegin}{\DIFOdelbegin \let\includegraphics\DIFdelincludegraphics} 
\DeclareRobustCommand{\DIFdelend}{\DIFOaddend \let\includegraphics\DIFOincludegraphics} 
\LetLtxMacro{\DIFOaddbeginFL}{\DIFaddbeginFL} 
\LetLtxMacro{\DIFOaddendFL}{\DIFaddendFL} 
\LetLtxMacro{\DIFOdelbeginFL}{\DIFdelbeginFL} 
\LetLtxMacro{\DIFOdelendFL}{\DIFdelendFL} 
\DeclareRobustCommand{\DIFaddbeginFL}{\DIFOaddbeginFL \let\includegraphics\DIFaddincludegraphics} 
\DeclareRobustCommand{\DIFaddendFL}{\DIFOaddendFL \let\includegraphics\DIFOincludegraphics} 
\DeclareRobustCommand{\DIFdelbeginFL}{\DIFOdelbeginFL \let\includegraphics\DIFdelincludegraphics} 
\DeclareRobustCommand{\DIFdelendFL}{\DIFOaddendFL \let\includegraphics\DIFOincludegraphics} 
\makeatletter 
\let\sout@orig\sout 
\renewcommand{\sout}[1]{\ifmmode\text{\sout@orig{\ensuremath{#1}}}\else\sout@orig{#1}\fi} 
\makeatother 
\RequirePackage{listings} 
\RequirePackage{color} 
\lstdefinelanguage{DIFcode}{ 
  moredelim=[il][\color{red}\sout]{\%DIF\ <\ }, 
  moredelim=[il][\color{blue}\uwave]{\%DIF\ >\ } 
} 
\lstdefinestyle{DIFverbatimstyle}{ 
	language=DIFcode, 
	basicstyle=\ttfamily, 
	columns=fullflexible, 
	keepspaces=true 
} 
\lstnewenvironment{DIFverbatim}{\lstset{style=DIFverbatimstyle}}{} 
\lstnewenvironment{DIFverbatim*}{\lstset{style=DIFverbatimstyle,showspaces=true}}{} 
\begin{document}

\title{Ground state exciton-polariton condensation by coherent Floquet driving}

\author{A. S. Kuznetsov}
\email[corresponding author: ]{kuznetsov@pdi-berlin.de}
\affiliation{Paul-Drude-Institut f{\"u}r Festk{\"o}rperelektronik, Leibniz-Institut im Forschungsverbund Berlin e.V., Hausvogteiplatz 5-7, 10117 Berlin, Germany}

\author{I. Carraro-Haddad}
\affiliation{Bariloche Atomic Centre and Balseiro Institute, National Council for Scientific and Technical Research, 8400 S.C. de Bariloche, R.N., Argentina}
\affiliation{Institute of Nanoscience and Nanotechnology, National Council for Scientific and Technical Research, 8400 Bariloche, Argentina}

\author{G. Usaj}
\affiliation{Bariloche Atomic Centre and Balseiro Institute, National Council for Scientific and Technical Research, 8400 S.C. de Bariloche, R.N., Argentina}
\affiliation{Institute of Nanoscience and Nanotechnology, National Council for Scientific and Technical Research, 8400 Bariloche, Argentina}

\author{K. Biermann}
\affiliation{Paul-Drude-Institut f{\"u}r Festk{\"o}rperelektronik, Leibniz-Institut im Forschungsverbund Berlin e.V., Hausvogteiplatz 5-7, 10117 Berlin, Germany}

\author{A. Fainstein}
\affiliation{Bariloche Atomic Centre and Balseiro Institute, National Council for Scientific and Technical Research, 8400 S.C. de Bariloche, R.N., Argentina}
\affiliation{Institute of Nanoscience and Nanotechnology, National Council for Scientific and Technical Research, 8400 Bariloche, Argentina}

\author{P. V. Santos}
\affiliation{Paul-Drude-Institut f{\"u}r Festk{\"o}rperelektronik, Leibniz-Institut im Forschungsverbund Berlin e.V., Hausvogteiplatz 5-7, 10117 Berlin, Germany}

\date{\today}
\begin{abstract}
The on-demand selective population transfer between states in multilevel quantum systems is a challenging problem with implications for a wide-range of physical platforms including photon and exciton-polariton Bose-Einstein condensates (BECs). Here, we introduce an universal strategy for this selective transfer based on a strong time-periodic energy modulation, which is experimentally demonstrated by using a GHz acoustic wave to control the gain and loss of confined modes of an exciton-polariton BEC in a microcavity. The harmonic acoustic field shifts the energy of the excitonic BEC component relative to the photonic ones, which generates a dynamic population transfer within a multimode BEC that can be controlled by the acoustic amplitude. In this way, the full BEC population can be selectively transferred to the ground state to yield a single-level emission consisting of  a spectral frequency comb with GHz repetition rates as well as picosecond-scale correlations. A theoretical model reproduces the observed time evolution and reveals a dynamical interplay between bosonic stimulation and the adiabatic Landau-Zener-like population transfer. Our approach provides a new avenue for the Floquet engineering of light-matter systems and enables tunable single- or multi-wavelength ultrafast pulsed laser-like emission for novel information technologies.
\end{abstract}

\maketitle 




\section{Introduction}
\label{Introduction}

Coherent and tunable population transfer between states in a multilevel system on a timescale comparable to its microscopic coherence and dynamics is a challenging problem appearing in several physical contexts including thermalization, lasing, and Bose-Einstein condensation (BEC). As an example, optically excited microcavity (MC) exciton-polaritons BECs~\cite{Weisbuch1992,Kasprzak2006} require an efficient process to transfer carriers from the pumped state to the final BEC state~\cite{Wouters2007}. In the presence of the spatial confinement, the condensation is typically a multimode one~\cite{Eastham2008,Grosso2014}. Generally, the condensation process under non-resonant excitation depends critically on the energy separation and spatial overlap between the pumped exciton-like reservoir and the confined states~\cite{Wouters2010}. In light of recent advances in photonic ground-state (GS) BECs~\cite{Schofield2024,Pieczarka2024}, it becomes important to realize control mechanisms enabling the on-demand, selective condensation at a particular level. To address this challenge, different schemes have been realized to dynamically modify the polariton BEC relaxation pathways using, for example, short optical pulses~\cite{Redondo2024}, Lotka-Volterra dynamics~\cite{Topfer2020}, as well as resonantly via OPO~\cite{Ferrier2010,Krizhanovskii2013,Kuznetsov2020b}.

Coherent time-periodic modulation of Hamiltonians is a powerful tool for Floquet engineering~\cite{Oka2019} of quantum systems~\cite{Silveri2017}. For example, periodically sweeping one energy level across the other allows to realize dynamic level crossing. In the case of a slow passage, the population follows the instantaneous eigenstate -- the adiabatic Landau-Zener transition~\cite{Ivakhnenko2023}. 
Recently a MHz-frequency modulation of a single-mode photonic BEC population in a quantum gas has been demonstrated, promising access to non-Hermitian topology~\cite{Erglis2025}. In solid state systems, the strain of piezoelectrically excited acoustic phonons can be used to modulate the energies of opto-electronic resonances~\cite{DeLima2005,Couto2009,Berstermann2009,Fuhrmann2011,Jusserand2015,Weiss2016,Hernandez-Minguez2020}. Very large energy modulation can be realized using cavity confinement and piezoelectrically~\cite{Sesin2023}, and impulsively~\cite{Berstermann2012} excited phonons.

\begin{figure*}[t]
	\centering
		\includegraphics[width=1\textwidth, keepaspectratio=true]{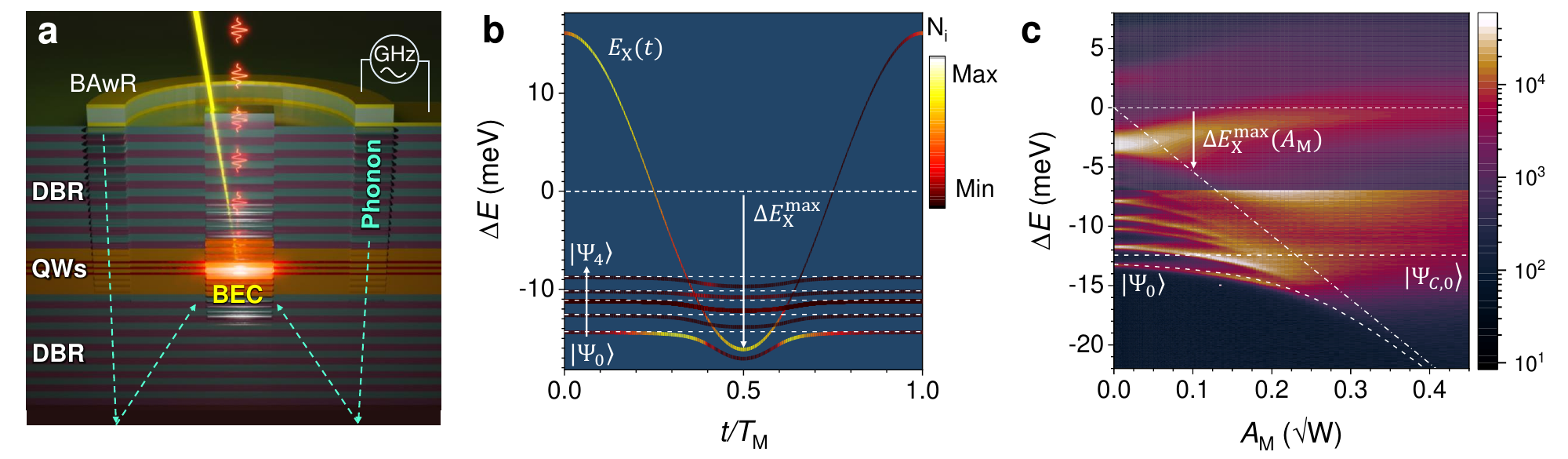}
		\caption{
            {\bf Acoustically induced dynamic level anti-crossings.}
            {\bf a} Schematic view of the device for the GHz acoustic control of microcavity polariton BECs based on a $\mu m$-sized intracavity trap. The harmonic strain field (phonon) is produced by driving the ring-shaped bulk acoustic resonator (BAwR) on the sample surface with a radio-frequency signal. The phonon is frequency-tuned to the cavity acoustic resonance and injected into the trap via reflections at the polished surfaces of the sample. The phonon modulates the energy and intensity of a non-resonantly excited (yellow beam) confined BEC, leading to the generation of coherent optical pulses (wiggly red lines) at GHz rates.
            {\bf b} Simulated periodic anti-crossing of the acoustically modulated bare exciton energy, $E_\mathrm{X}(t)$ (cosine-like curve), with multiple confined bare photon modes (horizontal white dashed lines) for a fixed modulation amplitude over one modulation period $T_\mathrm{M} = 2\pi/\Omega_\mathrm{M}$. The strong light-matter coupling results in polariton modes -- colored curves (only the lowest energy modes $|\Psi_{i}\rangle$ with  $i =0$ to 4 are shown). The line colors encode the population of the corresponding polariton and bare exciton states weighted by their photonic and excitonic fractions, respectively. The maximum bare exciton energy excursion is denoted by $\Delta E_\mathrm{X}^\mathrm{max}$. The energy scale is relative to the unmodulated bare exciton energy (horizontal dashed line at 0 meV).
            {\bf c} Experimental dependence of the trap photoluminescence (PL) on the acoustic modulation amplitude, $A_\mathrm{M}$, recorded for an optical excitation power below the BEC threshold. The diagonal dash-dotted line indicates the simulated maximum excursion of the bare exciton energy, $\Delta E_\mathrm{X}^\mathrm{max}(A_\mathrm{M})$, relative to the unmodulated exciton (horizontal dashed line at 0 meV). The curved dashed white line represents the polariton ground state energy, $|\Psi_{0}\rangle$, of the trap and the horizontal dashed white line its bare photonic energy, $|\Psi_\mathrm{C,0}\rangle$. For clarity, the intensity of the spectral range below -7~meV was multiplied by a factor of 50.         
            }
	\label{Fig1}
\end{figure*}

Here, we demonstrate the selective population transfer within a multilevel polariton BEC by creating a Floquet-like time-dependent Hamiltonian, $H(t) = H(t + T_\mathrm{M})$ via the harmonic energy modulation by a GHz acoustic strain with frequency $f_\mathrm{M} = 1/T_\mathrm{M}$. The studies were performed on a zero-dimensional polariton trap formed in a hybrid photon-phonon, patterned (Al,Ga)As MC~\cite{Fainstein2013, ElDaif2006}, as schematically illustrated in Fig.~\ref{Fig1}a.  For strong non-resonant optical pumping, the trap exhibits multimode BEC lasing at different confined states. The phonon strain, injected from a piezoelectric bulk acoustic wave resonator (BAwR)~\cite{Machado2019}, predominantly couples to the polariton excitonic levels~\cite{Sesin2023}. The electrical generation enables fine control of both the frequency and amplitude of the monochromatic strain for the dynamic modulation of the quantum well (QW) excitonic energies~\cite{DeLima2006}, as has been extensively addressed in previous studies using both surface~\cite{DeLima2006,Cerda-Mendez2010} and bulk acoustic waves~\cite{Kuznetsov2021}.

In contrast to previous studies, we induce multiple time-periodic avoided crossings between the bare exciton level and the bare confined photonic states using a large-amplitude harmonic acoustic modulation, as is illustrated in Fig.~\ref{Fig1}b. The energy detuning-dependent dynamics, then lead to very fast changes of the populations of the confined BEC modes. We precisely tune the maximum excursion of the bare exciton energy by the amplitude of the acoustic wave to match the static detuning between the exciton and a bare photonic mode, and demonstrate a selective population transfer to the mode. Essentially, full selectivity is achieved when the exciton energy approaches the photonic ground state, leading to a strong single-mode emission. We study the coherence in the spectral and time domains to confirm Floquet character of the modulation and demonstrate coherent pulsed emission. We present a detailed theoretical model, which explains the transfer and pulsed emission as an interplay between the stimulated scattering and 
the adiabatic Landau-Zener-like transfer of population.

%


\section{Results}
\label{Results}

The acousto-polaritonic device studied here (cf. Fig.~\ref{Fig1}a) is based on a patterned hybrid phonon-photon AlGaAs MC with multiple GaAs quantum wells (QWs) placed within the spacer region between two distributed Bragg reflectors (DBRs)~\cite{Fainstein2013}. The DBRs and the spacer are engineered to  maximize the overlap between the strain field of electrically excited  7~GHz phonon and the optical 370~THz ($\sim$1500 meV) photon modes at the QW positions. Both photonic and phononic states are vertically and laterally confined within a few-$\mu$m-wide thicker region of the MC spacer (simply, a trap). At 10~K, the strong-coupling between  QW excitons and the confined photon states leads to the formation of zero-dimensional polariton states~\cite{Kuznetsov2018}. We present results for a $4 \times 4 ~\mu m^2$ trap with photon-exciton detuning of approx. $-13$~meV. A ring-shaped bulk acoustic wave resonator (BAwR) fabricated around the trap excites a $f_\mathrm{M}=7$~GHz acoustic strain field polarized along the cavity growth axis. The strain couples predominantly to the excitonic polariton component via the deformation potential leading to peak energy modulation amplitudes that can exceed 25~meV~\cite{Sesin2023}. A detailed description of the MC sample can be found in Methods. 

\begin{figure*}[t]
	\centering
		\includegraphics[width=1.0\textwidth, keepaspectratio=true]{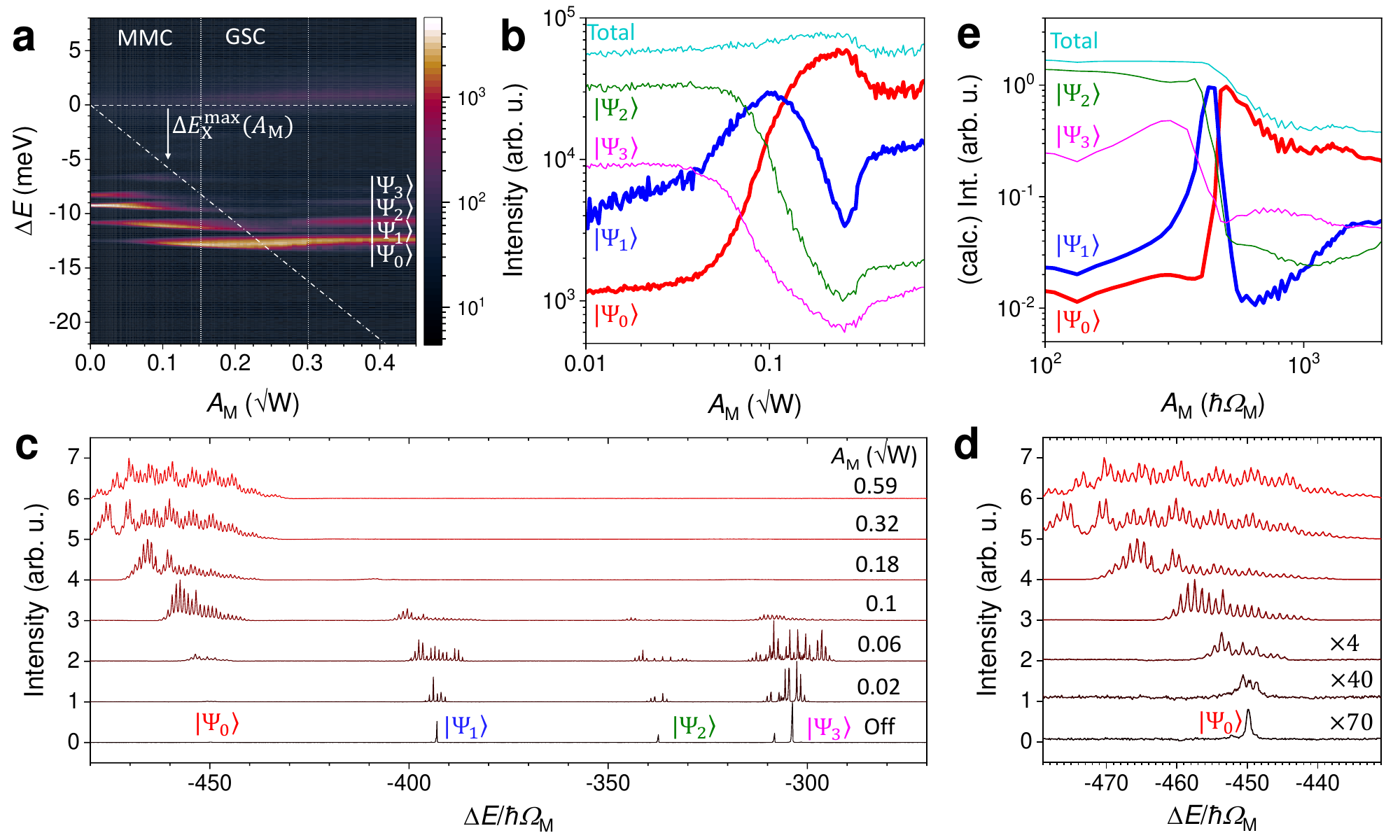}
		\caption{
            {\bf Modulation-induced population transfer and coherence.}
            {\bf a} Experimental dependence of the trap PL on the acoustic modulation amplitude, $A_\mathrm{M}$, recorded for an optical excitation power $P_\mathrm{exc} \approx 1.7 P_\mathrm{th}$ above the BEC threshold ($P_\mathrm{th}$). The lowest energy BEC modes are denoted $|\Psi_{i}\rangle$. The regimes of non-equilibrium lasing (MMC) and single-mode ground state (GS) condensation (GSC) are indicated. The diagonal dash-dotted line indicates the maximum excursion of the bare exciton energy, $\Delta E_\mathrm{X}^\mathrm{max}(A_\mathrm{M})$, obtained from Fig.~\ref{Fig1}c. The energy scale is referenced to the unmodulated exciton (horizontal dashed line at 0~meV).
            {\bf b} Experimental time-integrated PL of the polariton states together with the total PL (labeled in the panel) determined from Fig.~\ref{Fig2}a as a function of $A_\mathrm{M}$.
            {\bf c} High-resolution PL spectra of the trap for several $A_\mathrm{M}$, recorded under conditions similar to Fig.~\ref{Fig2}a. The spectra are normalized in amplitude and vertically shifted for clarity. The energy scale is relative to the bare exciton energy (cf.~panels Fig.~\ref{Fig2}a) and normalized to the modulation quantum $\hbar \Omega_\mathrm{M} = 28.7~\mu$eV. The sharp comb peaks are phonon sidebands displaced by $\hbar \Omega_\mathrm{M}$ resulting from the modulation of the GS ($i=0$) and of the first three excited states ($i=1,2,3$), $|\Psi_{i}\rangle$.
            {\bf d} Same as {\bf c}, but expanded over an energy range around GS centered at $\Delta E/(\hbar \Omega_\mathrm{M})=-449$.
            {\bf e} Simulated PL intensities as a function of $A_\mathrm{M}$ for the experimental conditions in panel {\bf a}.
            }
	\label{Fig2}
\end{figure*}

Figure~\ref{Fig1}b shows the time evolution of the confined states as calculated by the model discussed later. For clarity, only the bare exciton and the lowest five confined photonic modes are shown. The reference energy for the vertical axis corresponds to the bare unstrained exciton energy, $E_{X}^{*}$. The latter is harmonically modulated by the strain, i.e., $E_\mathrm{X}(t) = E_{X}^{*} + \Delta E_\mathrm{X}(t) =  E_{X}^{*} + E_\mathrm{M}\cos(\Omega_\mathrm{M} t)$, where $\Omega_\mathrm{M} = 2 \pi f_\mathrm{M}$. The energy modulation amplitude, $E_\mathrm{M}$, is proportional to the acoustic strain amplitude. At zero time, the exciton is well above the photonic states: $\Delta E_\mathrm{X}(0) \gg \hbar \Omega_\mathrm{R}$, where $\hbar \Omega_\mathrm{R}$ is the light-matter Rabi-coupling. Hence, the lower-energy polariton modes are photon-like. With increasing tensile strain, the bare exciton red-shifts and consecutively anti-crosses with multiple confined photonic modes. During the second half of the acoustic cycle, QWs are compressed and the bare exciton shift is reversed.

Figure~\ref{Fig1}c shows the dependence of the time-integrated photoluminescence (PL) spectrum of a trap under an excitation power below the BEC threshold ($P_\mathrm{exc}<P_\mathrm{th}$) on the acoustic amplitude ($A_\mathrm{M}$). The latter is proportional to the square-root of the nominal rf power applied to the BAwR at $\sim 7$~GHz frequency. For $A_\mathrm{M} = 0$, the emission is dominated by the PL line red-shifted by 3~meV corresponding to the trap barrier, which is also a polariton state~\cite{Kuznetsov2018}. The weaker lines at lower energies are the confined polariton modes. Increasing $A_\mathrm{M}$ broadens the levels with the $|\Psi_\mathrm{0}\rangle$ broadening exceeding 10~meV for $A_\mathrm{M} > 0.4$. In the time-integrated spectra, this broadening arises from the time-evolution of the states illustrated in Fig.~\ref{Fig1}b. Using the procedure described in Supplementary Information (SI-Sec.2), we show that the bare exciton crosses the bare photonic ground state $|\Psi_\mathrm{C,0}\rangle$ for $A_\mathrm{M} \approx 0.25$, as indicated by the dashed lines in Fig.~\ref{Fig1}c.

We now consider the acoustic modulation in the BEC regime, i.e., $P_\mathrm{exc}>P_\mathrm{th}$, which is depicted in Fig.~\ref{Fig2}a. In the absence of acoustic excitation, we observe a multi-mode condensation (MMC)  where the confined polariton modes compete for the gain~\cite{Topfer2020}. The emission of $|\Psi_{0}\rangle$ is about 30 times smaller than $|\Psi_{2}\rangle$. The excitation power dependence of the PL is discussed in SI-Sec.3.

The first striking observation from Fig.~\ref{Fig2}a is that as $A_\mathrm{M}$ increases and the levels broaden in energy, the overall PL intensity gradually shifts towards $|\Psi_{0}\rangle$. The dependence of the integrated PL intensities from confined levels on $A_\mathrm{M}$ is summarized in Fig.~\ref{Fig2}b. The higher excited states $|\Psi_{2}\rangle$ and $|\Psi_{3}\rangle$, which dominate the emission at low $A_\mathrm{M}$, start to be depleted when $A_\mathrm{M} \approx 0.05$ and $A_\mathrm{M} \approx 0.065$, respectively. In contrast, the PL intensities from $|\Psi_{1}\rangle$ and $|\Psi_{0}\rangle$ increase until they reach a maximum at $A_\mathrm{M} \approx 0.1$ and $A_\mathrm{M} \approx 0.25$, respectively.

The second remarkable observation is the onset of an essentially single-mode condensation (SMC) regime when the maximum bare exciton red-shift (diagonal dash-dot line) approaches $|\Psi_{0}\rangle$ (i.e., in the range $0.15 < A_\mathrm{M} < 0.3$ in Figs.~\ref{Fig2}a and \ref{Fig2}b). In this regime, PL is dominated by $|\Psi_{0}\rangle$. For $A_\mathrm{M} \approx 0.25$, the $|\Psi_{0}\rangle$ PL exceeds the one from the other levels by at least an order of magnitude, thus proving a very selective particle transfer directly to the ground state. When $A_\mathrm{M} > 0.35$, there is a partial recovery of MMC with level intensity and spectral profiles weakly dependent on $A_\mathrm{M}$. A similar evolution of PL with $A_\mathrm{M}$ has also been observed for other excitation powers (cf. SI-Sec.4). 

The total PL intensity integrated over all levels (light-blue line in Fig.~\ref{Fig2}b) remains approximately constant over the whole $A_\mathrm{M}$ range. Hence, the acoustic modulation essentially redistributes polaritons among the levels. That is, tunable acoustic amplitude allows to control the condensation pathway to a particular level. 

\begin{figure}[tbhp]
	\centering
		\includegraphics[width=1.0\columnwidth, keepaspectratio=true]{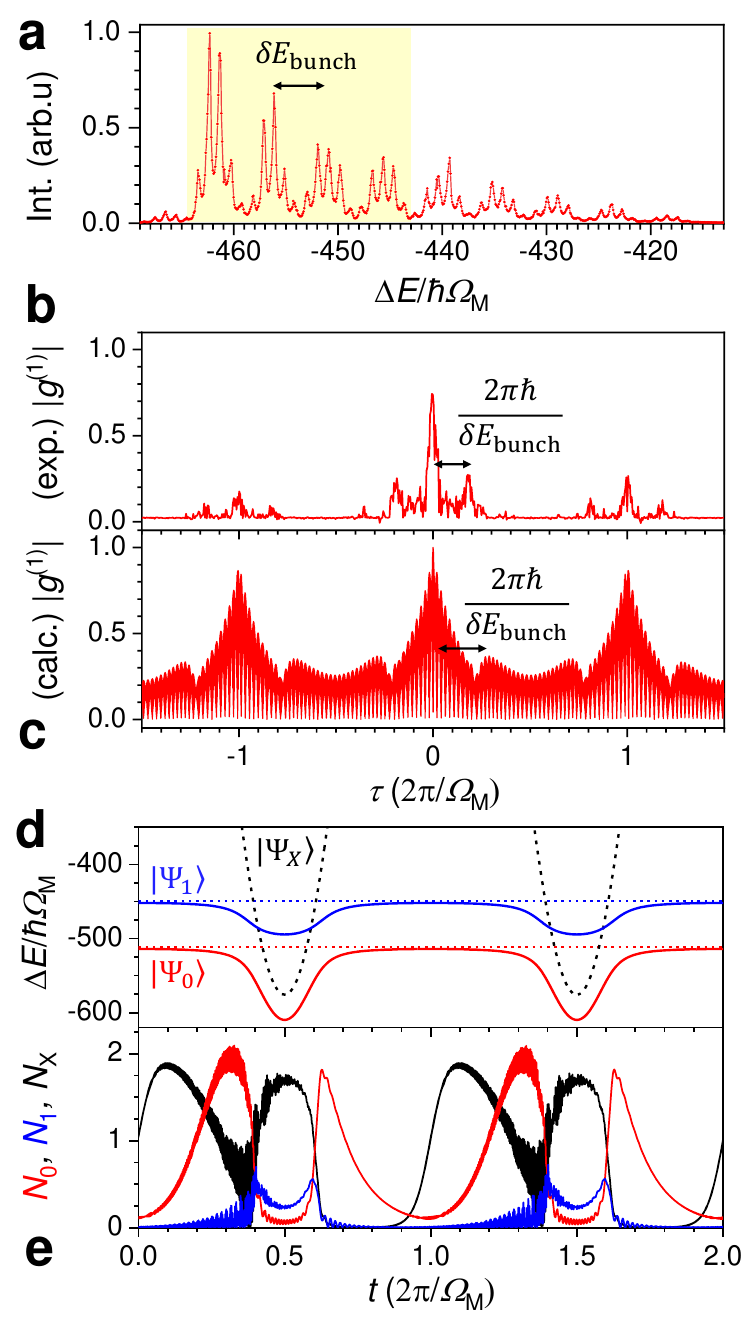}
		\caption{
            {\bf BEC pulsing under the acoustic modulation.} 
            {\bf a} High-resolution PL spectrum of the BEC GS ($|\Psi_{0}\rangle$) recorded for $A_\mathrm{M} = 0.59~\sqrt{W}$. The energy scale is referenced to the bare exciton energy and normalized to the modulation quantum $\hbar\Omega_\mathrm{M}$. The sideband bunching period is denoted $\delta E_\mathrm{bunch}$.
            {\bf b} First order auto-correlation function for the GS, $|g^{(1)}(\tau)|$ as a function of the delay $\tau$, as determined from the visibility of the $|g^{(1)}(\tau)|$ fringes in the energy range marked in panel {\bf a} by the yellow rectangle, 
            and {\bf c} the calculated $|g^{(1)}(\tau)|$.
            {\bf d} Simulated time-evolution of the energies of the bare exciton ($|\Psi_\mathrm{X}\rangle$, black) and polariton GS ($|\Psi_{0}\rangle$, red) and first excited ($|\Psi_{1}\rangle$, blue) states in the steady-state regime. The dashed horizontal lines show the energies of the bare photonic levels.
            {\bf e} Corresponding evolutions of the steady-state level populations of the bare exciton ($N_\mathrm{X}$, black) as well as of the ground ($N_\mathrm{0}$, red) and first excited ($N_\mathrm{1}$, blue) bare photon states for the modulation conditions in panel {\bf d}. The $N_{1}$ population has been scaled by a factor of 10 for better visibility. 
            }

	\label{Fig3}
\end{figure}


Despite the large energy modulation amplitudes, the high temporal coherence of the confined BECs is maintained over a wide $A_{M}$ range, as demonstrated by the high-resolution PL spectra of Fig.~\ref{Fig2}c, which were recorded under conditions similar to Fig.~\ref{Fig2}a. In the absence of the acoustic field, the PL consists of spectrally narrow peaks (linewidth $\gamma_\mathrm{BEC} < \Omega_\mathrm{M}$) from BECs at multiple excited levels. The modulation induces the formation of a frequency comb of resolved-sidebands separated by $\hbar \Omega_\mathrm{M}$ around the unperturbed BECs. The number of sidebands in each $|\Psi_{i}\rangle$ comb increases with $A_\mathrm{M}$. For small $A_\mathrm{M}$, the intensity of sidebands follows a Bessel distribution~\cite{Kuznetsov2023}, which is symmetric relative to the unmodulated resonance. For moderate and high $A_\mathrm{M}$, the comb spectral shape is strongly asymmetric. At the intermediate $A_\mathrm{M}$, several frequency combs are simultaneously present indicating the Floquet nature of the modulation. The GS, $|\Psi_{0}\rangle$, emission is very weak for $A_\mathrm{M} < 0.06$ but dominates for $A_\mathrm{M} > 0.1$. 
As detailed in SI-Sec.5, the spectral width of the individual lines of $|\Psi_{0}\rangle$ comb in Fig.~\ref{Fig2}d initially decreases with $A_\mathrm{M}$ from $\sim 0.7 \hbar \Omega_\mathrm{M} (A_\mathrm{M} = 0)$ to $\sim 0.3 \hbar \Omega_\mathrm{M} (A_\mathrm{M} = 0)$ and then increases again to $\sim 0.7 \hbar \Omega_\mathrm{M}$ at $A_\mathrm{M} = 0.59$. This behavior correlates with the increase of the $|\Psi_{0}\rangle$ intensity followed by its reduction, which can be due to interactions with the continuum of electronic states above the bare exciton energy.

We now address the time dynamics of the polariton system via studies of the first-order autocorrelation function ($|g^{(1)}|$) as a function of time delay ($\tau$) using an interferometric setup (see SI-Sec.6). The experiments were performed under a strong acoustic modulation ($A_\mathrm{M}=0.56$) and $P_\mathrm{exc} \approx 4P_\mathrm{th}$ yielding the PL spectrum for the GS displayed in Fig.~\ref{Fig3}a.  In addition to the modulation frequency comb, the sidebands show a clear bunching with a spacing $\delta E_\mathrm{bunch} \approx 5 \hbar \Omega_\mathrm{M}$. The corresponding $|g^{(1)}(\tau)|$ profile is displayed in Fig.~\ref{Fig3}b obtained for the energy range indicated in Fig.~\ref{Fig3}a by the yellow background. $|g^{(1)}(\tau)|$ shows clear peaks at $\pm T_\mathrm{M} = 2\pi / \Omega_\mathrm{M}$, thus demonstrating that the $|\Psi_{0}\rangle$ coherence persists over several acoustic modulation periods. In contrast, the modulation sidebands are not resolved for the excited state, which is weakly populated: accordingly, its $|g^{(1)}(\tau)|$ only shows a single peak at zero time delay, see SI-Sec.7. Furthermore, the GS profile has satellites at delays $\tau_\mathrm{sat} = \pm0.2 \times T_\mathrm{M}$ around the main peaks at $0, \pm 1 \times T_\mathrm{M}$, which correlate with the bunching of sidebands in Fig.~\ref{Fig3}a, i.e., $|\tau_\mathrm{sat}| = 2 \pi \hbar/\delta E_\mathrm{bunch}$. The narrow temporal width, triangular shape of $|g^{(1)}|$ peaks and bunching provide an indirect evidence of the pulsed character of the emission, as will be elucidated by the model below.

A theoretical model was developed based on the following three main assumptions (see SI-Sec.8 for in-depth discussion). (i) The strain field only modulates the  bare exciton state, which is populated by the non-resonant optical excitation. (ii)  The exciton feeds the polariton modes through both the stimulated scattering and the resonant injection due to the strong light-matter coupling. The latter, under the modulation, gives rise to a very efficient adiabatic Landau-Zener-like transfer. (iii) When the exciton energy is shifted below the corresponding bare photonic state, the stimulated scattering is suppressed since the relaxation from the exciton to the confined states becomes significantly less probable and, at the same time, the losses are increased due to the overlap with the continuum of higher-energy exciton states. These processes are incorporated into an effective Gross-Pitaevskii Hamiltonian describing the coupling between exciton and photon modes given by:

\begin{widetext}
\begin{eqnarray}
i\hbar \frac{d\tilde{\psi}_\mathrm{X}}{dt} &=& \left[ \varepsilon_\mathrm{X} + A_\mathrm{M} \cos{\left(\Omega_\mathrm{M} t \right)}\right] \tilde{\psi}_\mathrm{X} - \sum^{n}_{j=1} J_j \tilde{\psi}_{\mathrm{C},j} + \frac{i\hbar \gamma_\mathrm{X}}{2}  \left( \tilde{P} - \alpha_\mathrm{X} \tilde{n}_\mathrm{X} - \sum^{n}_{j=1} \alpha_j \tilde{n}_{\mathrm{C},j}\right) \tilde{\psi}_\mathrm{X}\,, \label{GPX}\\
%
i\hbar \frac{d\tilde{\psi}_{\mathrm{C},j}}{dt} &=& \varepsilon_{\mathrm{C},j} \tilde{\psi}_{\mathrm{C},j} - J_j \tilde{\psi}_\mathrm{X} + \frac{i\hbar}{2}  \left( \gamma_\mathrm{X} \alpha_{j} \tilde{n}_\mathrm{X} - \gamma_{\mathrm{C},j}\right) \tilde{\psi}_{\mathrm{C},j}\,.     \label{GPC}
\end{eqnarray}
\end{widetext}

\noindent Equation (\ref{GPX}) models the dynamics of the bare exciton state, $\psi_\mathrm{X}$. The first term on the rhs represents the exciton energy, $\varepsilon_\mathrm{X}$, modulated by the time-dependent strain. The second term accounts for the coherent light-matter coupling between the exciton and the bare confined photon states, $\psi_{\mathrm{C},j}$ ($j = 0,1,\dots,n-1)$, where $n$ is the total number of confined photonic states, leading to the formation of polaritons. The strength of this coupling is characterized by the Rabi coupling energy $J_j = \hbar \Omega_{\mathrm{R},j}$, which governs the rate of population exchange between the exciton and each confined state. Finally, the third term accounts for the non-Hermitian nature of the system in the BEC regime. Here, $\tilde{P}$ represents the net injection rate of particles into the exciton state state (relative to the level at the BEC threshold) from the optically non-resonantly pumped reservoir, while $\gamma_\mathrm{X}$ denotes the exciton decay rate. The term $\alpha_\mathrm{X} \tilde{n}_X$ describes exciton saturation effects
, where $\alpha_\mathrm{X}$ is the filling rate constant, and $\tilde{n}_\mathrm{X} = \left|\tilde{\psi}_\mathrm{X}\right|^2$ is the exciton density. Finally, $\sum^{n}_{j=1} \alpha_j \tilde{n}_{\mathrm{C},j}$ represents the total stimulated scattering rate into the confined photonic states, where $\alpha_j$ is the filling rate constant for each state, and $\tilde{n}_{\mathrm{C},j} = \left|\tilde{\psi}_{\mathrm{C},j}\right|^2$ is the corresponding density.

Similarly, Eq.~(\ref{GPC}) describes the dynamics of the confined light modes. The first term on the rhs corresponds to their bare energies  $\varepsilon_{\mathrm{C},j}$, which depend on the lateral confinement potential. The second term accounts for the light-matter interaction with the exciton state. The third term describes the filling of the confined states through stimulated scattering from the exciton at a rate $\gamma \alpha_j \tilde{n}_\mathrm{X}$ as well as their dissipation due to cavity losses with a decay rate $\gamma_j$. The losses depend on the detuning of between the exciton and photonic modes and, thus, on time, as discussed in SI-Sec.8C. The parameters used in the simulations are listed in SI-Sec.8D.

The dependence of time-averaged populations of the different levels as predicted by the model is shown in Fig.~\ref{Fig2}e. The predictions excellently reproduce the progressive loading of the levels shown in Fig.~\ref{Fig2}b. 

Figure~\ref{Fig3}d,e shows, respectively, the simulated time-evolution of the energies and populations of the bare exciton ($\tilde{n}_\mathrm{X} =|\langle\psi_\mathrm{X}\rangle|^2$), ground ($\tilde{n_0} = |\langle\psi_{\mathrm{C},0}\rangle|^2$) and the first excited ($\tilde{n_1}=|\langle\psi_{\mathrm{C},1}\rangle|^2$) polaritons states in the steady-state regime. The short-time transient after the modulation turn-on is discussed in SI-Sec.9. The configuration of energy levels as well as the optical and acoustic excitation correspond approximately to the conditions of Fig.~\ref{Fig2}a for $A_\mathrm{M} \approx 0.25$. The population data of Fig.~\ref{Fig3}e is also represented in an alternative way by the color lines in Fig.~\ref{Fig1}b.

We consider the evolution of the populations  during the modulation cycle of Fig.~\ref{Fig3}e. For large negative dynamic detuning between the bare exciton and the photonic GS, $\delta_{\mathrm{C}_{0}-\mathrm{X}}(t = 0) \ll 0$, the exciton holds most of the population and photonic levels are essentially empty. As $\delta_{\mathrm{C}_{0}-\mathrm{X}}(t)$ reduces, the higher levels become progressively more populated, while the exciton is de-populated. For each level, the maximum population is achieved very close to the respective avoided crossing point, where the population transfer is due to the coherent Rabi oscillations.

As the exciton moves below a photonic level $|\Psi_{\mathrm{C},j}\rangle$, this level becomes depopulated. Below $|\Psi_{\mathrm{C},0}\rangle$, the stimulated scattering mechanism is suppressed and there is a partial recovery of the exciton population. As the exciton approaches $|\Psi_{\mathrm{C},0}\rangle$ from below, there is a fast Rabi transfer of the population, which leads to the second emission pulse at $t = 0.65 (2\pi/\Omega_\mathrm{M})$ in Fig.~\ref{Fig3}e. The time separation between the sub-pulses depends on the modulation amplitude. For $|\Psi_{C,0}\rangle$ crossing conditions, the sub-pulsing frequency is approx. $2\times\Omega_\mathrm{M}/2\pi$.

The $|\Psi_{C,0}\rangle$ time evolution of Fig.~\ref{Fig3}e leads to the $|g^{(1)}|$ shown in Fig.~\ref{Fig3}c (the corresponding comb spectrum is discussed in SI-Sec.10). The triangular shape of the pulses gives rise to the triangular shape of $|g^{(1)}|$, while the pulsing leads to the additional peaks at $\pm T_\mathrm{M}=2\pi/\Omega_\mathrm{M}$ and the features displaced by $2 \pi \hbar/\delta E_\mathrm{bunch}$ associated with the sub-pulsing -- the cause of the spectral bunching of sidebands. 


\section{Discussion}
\label{Discussions}

Our experiments demonstrate that a GHz-frequency acoustic modulation
can trigger single-mode, highly coherent BEC condensation in the ground state of a multi-level polariton system. The modulation allows to periodically move the (non-resonantly populated) exciton state through multiple avoided-crossings with confined photonic levels. This dynamic process enables selective seeding of the (initially weakly populated) GS with enough polaritons to trigger the stimulated scattering leading to persistent GS BEC, which then dominates the emission of the multi-level system. The coherent Floquet nature of the modulation is confirmed by high-resolution spectroscopy, which shows that the BEC emission consists of combs of sidebands separated by the modulation quantum. Time-domain auto-correlation studies reveal that the GS emission consists of a  train of sub-50~ps pulses with the GHz repetition rate. The results are quantitatively accounted for by a theoretical model, which takes into account both the dynamical stimulated scattering and the Rabi coupling-mediated transfer between the states. In this dynamic setting, the latter leads to the adiabatic Landau-Zener-like transitions.

A remarkable feature of the acoustically modulated GS BEC is that its temporal coherence remains longer than the modulation period  despite the very large modulation amplitudes, which can exceeded the light-matter coupling energy by over an order of magnitude. Counterintuitively, the long coherence persists even though the  GS instantaneous population can become fully depleted during a fraction of the modulation cycle. We argue that the modulation essentially transfers particles between the different states, thereby keeping the total particle number (and, thus, the coherence) constant. We envision that the same mechanism can also be applied to seed BECs in excited levels, as long as the interlevel splitting is comparable to or larger than the Rabi coupling ($\hbar \Omega_\mathrm{R}$).

Here, we modulate the system deep in the adiabatic regime. In the other limit, when $\Omega_\mathrm{R}^2 \leq \hbar \it{v}$, where $\it{v}$ is the energy modulation rate $v = dE/dt \sim A_\mathrm{M}$, the transition may become diabatic leading to Landau-Zener–Stückelberg–Majorana interferences~\cite{Ivakhnenko2023}.


Furthermore, the results open exciting prospects to coherently modulate BECs on lattices, which are a prerequisite for Floquet topological insulators~\cite{Oka2019}. The demonstrated modulation, when applied to lattices can be used to break the time-reversal symmetry to mimic  effective magnetic fields for photons to overcome the Lorentz reciprocity and enable the realization of non-reciprocal electromagnetic transmission on a chip. Indeed, the possibility of GHz modulation using an optical scheme was demonstrated recently~\cite{Redondo2024}. The realization of the spatio-temporal modulation at the lattice level would open the path to the demonstration of novel non-reciprocal polariton and photon propagation phenomena with potential applications in on-chip generation of coherent ps-pulses, control of photon BECs, nano-laser arrays, bidirectional microwave-to-optical signal conversion, and for simulators of complex classical and quantum physical problems.


\section*{Methods}

\textbf{Microcavity sample.}

(Al,Ga)As systems exhibit ``double magic coincidence''~\cite{Trigo2002,Fainstein2013} between light and GHz acoustic vibrations, which ensures that a (Al,Ga)As MC for near-infrared photons simultaneously confines GHz phonons. The details of a hybrid photon-phonon MC with polariton traps studied here were recently reported in Ref.~\cite{Kuznetsov2023}. The MC has a quality factor $Q \approx 5000$ and contains six 15-nm-thick GaAs quantum wells (QWs), leading to the light-matter Rabi coupling of $\sim 3$~meV. Importantly, the spacer of the cavity confines $\sim 7$~GHz longitudinal bulk acoustic wave (BAW), which has optimized coupling to the excitonic resonances via the deformation potential mechanism~\cite{Kuznetsov2021}.

Polaritons and phonons are both subject to lateral confinement created by the etch-and-overgrowth method~\cite{ElDaif2006}. The confinement is provided by creating a few-nm-thick and a few-$\mu$m-wide mesa in the spacer of the MC (but away from QWs)~\cite{Kuznetsov2018}. Due to the ``double magic coincidence'', the traps confine photons and phonons. Presented experiments are recoded on a nominally square $4 \times 4~\mu m^2$ trap for a detuning of approx. $-13$~meV between bare photon and exciton energies in the non-etched region. Further information about spatial dispersion of polariton modes and cavity light-matter properties is presented in SI-Sec.1. 

\textbf{Bulk acoustic wave transducers.}

The acoustic mode of $F_\mathrm{M} \approx 7$~GHz of the MC was resonantly driven by a bulk acoustic resonator (BAwR)~\cite{Machado2019} fabricated on top of the region containing the trap and excited using a radio-frequency generator, cf. Fig.~\ref{Fig1}a, described in the previous section. The wave is excited via the piezoelectric effect by the BAwR and away from the trap, and reaches the trap after lateral propagation involving multiple reflections between the polished bottom and top sides of the sample. The amplitude of the BAW can be precisely controlled by the radio-frequency power applied to BAwR. A radio-frequency probe was used to connect the bottom and top electrodes of the transducer to a radio-frequency generator with high-power output

\textbf{Optical measurements with acoustic modulation.}

Optical measurements under acoustic excitation were carried out in a cold-finger liquid-He cryogenic probe station at 10~K temperature. A single-mode continuous wave stabilized Ti-Sapphire laser was used to non-resonantly excite polaritons. Excitation wavelength was set to around 760 nm -- strongly non-resonant with respect to polariton modes around 810 nm. The laser was focused on the sample at normal incidence using a 10x objective resulting in a few-$\mu$m Gaussian-like spot positioned on the trap.

Standard PL measurements with moderate spectral resolution of 0.1~meV, were carried out by transferring PL image of the trap on the entrance slit of a single grating spectrometer and recorded using a nitrogen-cooled CCD camera. PL measurements with high spectral resolution of $\sim 1.2~\mu$eV (0.3~GHz) were performed using a voltage-tunable Fabry-Perot etalon~\cite{Kuznetsov2023}. Here, the PL was sent through a single mode fiber into the etalon, which acts as a high-resolution filter, and then coarsely resolved with another single grating spectrometer (0.1~meV resolution) and recorded with a nitrogen-cooled CCD camera. To avoid thermal drifts, the etalon was actively temperature-stabilized with an external heater. Long-pass filters were used to block the light from the pump laser.

Auto-correlation measurements were carried out using a modified Michelson interferometer with one arm of a fixed length and the other arm with a movable retro-reflector, forming a motorized delay line. The light was sent into the interferometer through an optical fiber. Half of the incident PL is sent through the fixed arm and then interfered on the CCD with the half that is delayed by the second arm. The relative difference between the length of the two arms translates to a temporal delay. The auto-correlation curves were then extracted from interferograms recorded by scanning the delay. The schematics of the setup and an example of a recorded interferogram are presented in SI.6. 

\section*{Data availability}
The measurement and numerical simulation data that support the findings within this study are included within the main text and Supplementary Information and can also be made available upon request from the corresponding author.

\bibliography{Mendeley_Bib}

\begin{thebibliography}{35}%
\makeatletter
\providecommand \@ifxundefined [1]{%
 \@ifx{#1\undefined}
}%
\providecommand \@ifnum [1]{%
 \ifnum #1\expandafter \@firstoftwo
 \else \expandafter \@secondoftwo
 \fi
}%
\providecommand \@ifx [1]{%
 \ifx #1\expandafter \@firstoftwo
 \else \expandafter \@secondoftwo
 \fi
}%
\providecommand \natexlab [1]{#1}%
\providecommand \enquote  [1]{``#1''}%
\providecommand \bibnamefont  [1]{#1}%
\providecommand \bibfnamefont [1]{#1}%
\providecommand \citenamefont [1]{#1}%
\providecommand \href@noop [0]{\@secondoftwo}%
\providecommand \href [0]{\begingroup \@sanitize@url \@href}%
\providecommand \@href[1]{\@@startlink{#1}\@@href}%
\providecommand \@@href[1]{\endgroup#1\@@endlink}%
\providecommand \@sanitize@url [0]{\catcode `\\12\catcode `\$12\catcode `\&12\catcode `\#12\catcode `\^12\catcode `\_12\catcode `\%12\relax}%
\providecommand \@@startlink[1]{}%
\providecommand \@@endlink[0]{}%
\providecommand \url  [0]{\begingroup\@sanitize@url \@url }%
\providecommand \@url [1]{\endgroup\@href {#1}{\urlprefix }}%
\providecommand \urlprefix  [0]{URL }%
\providecommand \Eprint [0]{\href }%
\providecommand \doibase [0]{https://doi.org/}%
\providecommand \selectlanguage [0]{\@gobble}%
\providecommand \bibinfo  [0]{\@secondoftwo}%
\providecommand \bibfield  [0]{\@secondoftwo}%
\providecommand \translation [1]{[#1]}%
\providecommand \BibitemOpen [0]{}%
\providecommand \bibitemStop [0]{}%
\providecommand \bibitemNoStop [0]{.\EOS\space}%
\providecommand \EOS [0]{\spacefactor3000\relax}%
\providecommand \BibitemShut  [1]{\csname bibitem#1\endcsname}%
\let\auto@bib@innerbib\@empty
\bibitem [{\citenamefont {Weisbuch}\ \emph {et~al.}(1992)\citenamefont {Weisbuch}, \citenamefont {Nishioka}, \citenamefont {Ishikawa},\ and\ \citenamefont {Arakawa}}]{Weisbuch1992}%
  \BibitemOpen
  \bibfield  {author} {\bibinfo {author} {\bibfnamefont {C.}~\bibnamefont {Weisbuch}}, \bibinfo {author} {\bibfnamefont {M.}~\bibnamefont {Nishioka}}, \bibinfo {author} {\bibfnamefont {A.}~\bibnamefont {Ishikawa}},\ and\ \bibinfo {author} {\bibfnamefont {Y.}~\bibnamefont {Arakawa}},\ }\bibfield  {title} {\bibinfo {title} {{Observation of the coupled exciton-photon mode splitting in a semiconductor quantum microcavity}},\ }\href {https://doi.org/10.1103/PhysRevLett.69.3314} {\bibfield  {journal} {\bibinfo  {journal} {Phys. Rev. Lett.}\ }\textbf {\bibinfo {volume} {69}},\ \bibinfo {pages} {3314} (\bibinfo {year} {1992})}\BibitemShut {NoStop}%
\bibitem [{\citenamefont {Kasprzak}\ \emph {et~al.}(2006)\citenamefont {Kasprzak}, \citenamefont {Richard}, \citenamefont {Kundermann}, \citenamefont {Baas}, \citenamefont {Jeambrun}, \citenamefont {Keeling}, \citenamefont {Marchetti}, \citenamefont {Szyma{\'{n}}ska}, \citenamefont {Andr{\'{e}}}, \citenamefont {Staehli}, \citenamefont {Savona}, \citenamefont {Littlewood}, \citenamefont {Deveaud},\ and\ \citenamefont {Dang}}]{Kasprzak2006}%
  \BibitemOpen
  \bibfield  {author} {\bibinfo {author} {\bibfnamefont {J.}~\bibnamefont {Kasprzak}}, \bibinfo {author} {\bibfnamefont {M.}~\bibnamefont {Richard}}, \bibinfo {author} {\bibfnamefont {S.}~\bibnamefont {Kundermann}}, \bibinfo {author} {\bibfnamefont {A.}~\bibnamefont {Baas}}, \bibinfo {author} {\bibfnamefont {P.}~\bibnamefont {Jeambrun}}, \bibinfo {author} {\bibfnamefont {J.~M.~J.}\ \bibnamefont {Keeling}}, \bibinfo {author} {\bibfnamefont {F.~M.}\ \bibnamefont {Marchetti}}, \bibinfo {author} {\bibfnamefont {M.~H.}\ \bibnamefont {Szyma{\'{n}}ska}}, \bibinfo {author} {\bibfnamefont {R.}~\bibnamefont {Andr{\'{e}}}}, \bibinfo {author} {\bibfnamefont {J.~L.}\ \bibnamefont {Staehli}}, \bibinfo {author} {\bibfnamefont {V.}~\bibnamefont {Savona}}, \bibinfo {author} {\bibfnamefont {P.~B.}\ \bibnamefont {Littlewood}}, \bibinfo {author} {\bibfnamefont {B.}~\bibnamefont {Deveaud}},\ and\ \bibinfo {author} {\bibfnamefont {L.~S.}\ \bibnamefont {Dang}},\ }\bibfield  {title} {\bibinfo {title} {{Bose–Einstein condensation
  of exciton polaritons}},\ }\href {https://doi.org/10.1038/nature05131} {\bibfield  {journal} {\bibinfo  {journal} {Nature}\ }\textbf {\bibinfo {volume} {443}},\ \bibinfo {pages} {409} (\bibinfo {year} {2006})}\BibitemShut {NoStop}%
\bibitem [{\citenamefont {Wouters}\ and\ \citenamefont {Carusotto}(2007)}]{Wouters2007}%
  \BibitemOpen
  \bibfield  {author} {\bibinfo {author} {\bibfnamefont {M.}~\bibnamefont {Wouters}}\ and\ \bibinfo {author} {\bibfnamefont {I.}~\bibnamefont {Carusotto}},\ }\bibfield  {title} {\bibinfo {title} {{Excitations in a nonequilibrium bose-einstein condensate of exciton polaritons}},\ }\href {https://doi.org/10.1103/PhysRevLett.99.140402} {\bibfield  {journal} {\bibinfo  {journal} {Phys. Rev. Lett.}\ }\textbf {\bibinfo {volume} {99}},\ \bibinfo {pages} {140402} (\bibinfo {year} {2007})}\BibitemShut {NoStop}%
\bibitem [{\citenamefont {Eastham}(2008)}]{Eastham2008}%
  \BibitemOpen
  \bibfield  {author} {\bibinfo {author} {\bibfnamefont {P.~R.}\ \bibnamefont {Eastham}},\ }\bibfield  {title} {\bibinfo {title} {{Mode-locking and mode-competition in a non-equilibrium solid-state condensate}},\ }\href {https://doi.org/10.1103/PhysRevB.78.035319} {\bibfield  {journal} {\bibinfo  {journal} {Phys. Rev. B}\ }\textbf {\bibinfo {volume} {78}},\ \bibinfo {pages} {035319} (\bibinfo {year} {2008})}\BibitemShut {NoStop}%
\bibitem [{\citenamefont {Grosso}\ \emph {et~al.}(2014)\citenamefont {Grosso}, \citenamefont {Trebaol}, \citenamefont {Wouters}, \citenamefont {Morier-Genoud}, \citenamefont {Portella-Oberli},\ and\ \citenamefont {Deveaud}}]{Grosso2014}%
  \BibitemOpen
  \bibfield  {author} {\bibinfo {author} {\bibfnamefont {G.}~\bibnamefont {Grosso}}, \bibinfo {author} {\bibfnamefont {S.}~\bibnamefont {Trebaol}}, \bibinfo {author} {\bibfnamefont {M.}~\bibnamefont {Wouters}}, \bibinfo {author} {\bibfnamefont {F.}~\bibnamefont {Morier-Genoud}}, \bibinfo {author} {\bibfnamefont {M.~T.}\ \bibnamefont {Portella-Oberli}},\ and\ \bibinfo {author} {\bibfnamefont {B.}~\bibnamefont {Deveaud}},\ }\bibfield  {title} {\bibinfo {title} {{Nonlinear relaxation and selective polychromatic lasing of confined polaritons}},\ }\href {https://doi.org/10.1103/PhysRevB.90.045307} {\bibfield  {journal} {\bibinfo  {journal} {Phys. Rev. B - Condens. Matter Mater. Phys.}\ }\textbf {\bibinfo {volume} {90}},\ \bibinfo {pages} {045307} (\bibinfo {year} {2014})}\BibitemShut {NoStop}%
\bibitem [{\citenamefont {Wouters}\ \emph {et~al.}(2010)\citenamefont {Wouters}, \citenamefont {Liew},\ and\ \citenamefont {Savona}}]{Wouters2010}%
  \BibitemOpen
  \bibfield  {author} {\bibinfo {author} {\bibfnamefont {M.}~\bibnamefont {Wouters}}, \bibinfo {author} {\bibfnamefont {T.~C.}\ \bibnamefont {Liew}},\ and\ \bibinfo {author} {\bibfnamefont {V.}~\bibnamefont {Savona}},\ }\bibfield  {title} {\bibinfo {title} {{Energy relaxation in one-dimensional polariton condensates}},\ }\href {https://doi.org/10.1103/PhysRevB.82.245315} {\bibfield  {journal} {\bibinfo  {journal} {Phys. Rev. B - Condens. Matter Mater. Phys.}\ }\textbf {\bibinfo {volume} {82}},\ \bibinfo {pages} {245315} (\bibinfo {year} {2010})}\BibitemShut {NoStop}%
\bibitem [{\citenamefont {Schofield}\ \emph {et~al.}(2024)\citenamefont {Schofield}, \citenamefont {Fu}, \citenamefont {Clarke}, \citenamefont {Farrer}, \citenamefont {Trapalis}, \citenamefont {Dhar}, \citenamefont {Mukherjee}, \citenamefont {{Severs Millard}}, \citenamefont {Heffernan}, \citenamefont {Mintert}, \citenamefont {Nyman},\ and\ \citenamefont {Oulton}}]{Schofield2024}%
  \BibitemOpen
  \bibfield  {author} {\bibinfo {author} {\bibfnamefont {R.~C.}\ \bibnamefont {Schofield}}, \bibinfo {author} {\bibfnamefont {M.}~\bibnamefont {Fu}}, \bibinfo {author} {\bibfnamefont {E.}~\bibnamefont {Clarke}}, \bibinfo {author} {\bibfnamefont {I.}~\bibnamefont {Farrer}}, \bibinfo {author} {\bibfnamefont {A.}~\bibnamefont {Trapalis}}, \bibinfo {author} {\bibfnamefont {H.~S.}\ \bibnamefont {Dhar}}, \bibinfo {author} {\bibfnamefont {R.}~\bibnamefont {Mukherjee}}, \bibinfo {author} {\bibfnamefont {T.}~\bibnamefont {{Severs Millard}}}, \bibinfo {author} {\bibfnamefont {J.}~\bibnamefont {Heffernan}}, \bibinfo {author} {\bibfnamefont {F.}~\bibnamefont {Mintert}}, \bibinfo {author} {\bibfnamefont {R.~A.}\ \bibnamefont {Nyman}},\ and\ \bibinfo {author} {\bibfnamefont {R.~F.}\ \bibnamefont {Oulton}},\ }\bibfield  {title} {\bibinfo {title} {{Bose–Einstein condensation of light in a semiconductor quantum well microcavity}},\ }\href {https://doi.org/10.1038/s41566-024-01491-2} {\bibfield  {journal} {\bibinfo
  {journal} {Nat. Photonics}\ }\textbf {\bibinfo {volume} {18}},\ \bibinfo {pages} {1083} (\bibinfo {year} {2024})},\ \Eprint {https://arxiv.org/abs/2306.15314} {arXiv:2306.15314} \BibitemShut {NoStop}%
\bibitem [{\citenamefont {Pieczarka}\ \emph {et~al.}(2024)\citenamefont {Pieczarka}, \citenamefont {G{\c{e}}bski}, \citenamefont {Piasecka}, \citenamefont {Lott}, \citenamefont {Pelster}, \citenamefont {Wasiak},\ and\ \citenamefont {Czyszanowski}}]{Pieczarka2024}%
  \BibitemOpen
  \bibfield  {author} {\bibinfo {author} {\bibfnamefont {M.}~\bibnamefont {Pieczarka}}, \bibinfo {author} {\bibfnamefont {M.}~\bibnamefont {G{\c{e}}bski}}, \bibinfo {author} {\bibfnamefont {A.~N.}\ \bibnamefont {Piasecka}}, \bibinfo {author} {\bibfnamefont {J.~A.}\ \bibnamefont {Lott}}, \bibinfo {author} {\bibfnamefont {A.}~\bibnamefont {Pelster}}, \bibinfo {author} {\bibfnamefont {M.}~\bibnamefont {Wasiak}},\ and\ \bibinfo {author} {\bibfnamefont {T.}~\bibnamefont {Czyszanowski}},\ }\bibfield  {title} {\bibinfo {title} {{Bose–Einstein condensation of photons in a vertical-cavity surface-emitting laser}},\ }\href {https://doi.org/10.1038/s41566-024-01478-z} {\bibfield  {journal} {\bibinfo  {journal} {Nat. Photonics}\ }\textbf {\bibinfo {volume} {18}},\ \bibinfo {pages} {1090} (\bibinfo {year} {2024})}\BibitemShut {NoStop}%
\bibitem [{\citenamefont {{del Valle Inclan Redondo}}\ \emph {et~al.}(2024)\citenamefont {{del Valle Inclan Redondo}}, \citenamefont {Xu}, \citenamefont {Liew}, \citenamefont {Ostrovskaya}, \citenamefont {Stegmaier}, \citenamefont {Thomale}, \citenamefont {Schneider}, \citenamefont {Dam}, \citenamefont {Klembt}, \citenamefont {H{\"{o}}fling}, \citenamefont {Tarucha},\ and\ \citenamefont {Fraser}}]{Redondo2024}%
  \BibitemOpen
  \bibfield  {author} {\bibinfo {author} {\bibfnamefont {Y.}~\bibnamefont {{del Valle Inclan Redondo}}}, \bibinfo {author} {\bibfnamefont {X.}~\bibnamefont {Xu}}, \bibinfo {author} {\bibfnamefont {T.~C.}\ \bibnamefont {Liew}}, \bibinfo {author} {\bibfnamefont {E.~A.}\ \bibnamefont {Ostrovskaya}}, \bibinfo {author} {\bibfnamefont {A.}~\bibnamefont {Stegmaier}}, \bibinfo {author} {\bibfnamefont {R.}~\bibnamefont {Thomale}}, \bibinfo {author} {\bibfnamefont {C.}~\bibnamefont {Schneider}}, \bibinfo {author} {\bibfnamefont {S.}~\bibnamefont {Dam}}, \bibinfo {author} {\bibfnamefont {S.}~\bibnamefont {Klembt}}, \bibinfo {author} {\bibfnamefont {S.}~\bibnamefont {H{\"{o}}fling}}, \bibinfo {author} {\bibfnamefont {S.}~\bibnamefont {Tarucha}},\ and\ \bibinfo {author} {\bibfnamefont {M.~D.}\ \bibnamefont {Fraser}},\ }\bibfield  {title} {\bibinfo {title} {{Non-reciprocal band structures in an exciton–polariton Floquet optical lattice}},\ }\href {https://doi.org/10.1038/s41566-024-01424-z} {\bibfield  {journal}
  {\bibinfo  {journal} {Nat. Photonics}\ }\textbf {\bibinfo {volume} {18}},\ \bibinfo {pages} {548} (\bibinfo {year} {2024})}\BibitemShut {NoStop}%
\bibitem [{\citenamefont {T{\"{o}}pfer}\ \emph {et~al.}(2020)\citenamefont {T{\"{o}}pfer}, \citenamefont {Sigurdsson}, \citenamefont {Alyatkin},\ and\ \citenamefont {Lagoudakis}}]{Topfer2020}%
  \BibitemOpen
  \bibfield  {author} {\bibinfo {author} {\bibfnamefont {J.~D.}\ \bibnamefont {T{\"{o}}pfer}}, \bibinfo {author} {\bibfnamefont {H.}~\bibnamefont {Sigurdsson}}, \bibinfo {author} {\bibfnamefont {S.}~\bibnamefont {Alyatkin}},\ and\ \bibinfo {author} {\bibfnamefont {P.~G.}\ \bibnamefont {Lagoudakis}},\ }\bibfield  {title} {\bibinfo {title} {{Lotka-Volterra population dynamics in coherent and tunable oscillators of trapped polariton condensates}},\ }\href {https://doi.org/10.1103/PhysRevB.102.195428} {\bibfield  {journal} {\bibinfo  {journal} {Phys. Rev. B}\ }\textbf {\bibinfo {volume} {102}},\ \bibinfo {pages} {195428} (\bibinfo {year} {2020})}\BibitemShut {NoStop}%
\bibitem [{\citenamefont {Ferrier}\ \emph {et~al.}(2010)\citenamefont {Ferrier}, \citenamefont {Pigeon}, \citenamefont {Wertz}, \citenamefont {Bamba}, \citenamefont {Senellart}, \citenamefont {Sagnes}, \citenamefont {Lem{\`{a}}tre}, \citenamefont {Ciuti},\ and\ \citenamefont {Bloch}}]{Ferrier2010}%
  \BibitemOpen
  \bibfield  {author} {\bibinfo {author} {\bibfnamefont {L.}~\bibnamefont {Ferrier}}, \bibinfo {author} {\bibfnamefont {S.}~\bibnamefont {Pigeon}}, \bibinfo {author} {\bibfnamefont {E.}~\bibnamefont {Wertz}}, \bibinfo {author} {\bibfnamefont {M.}~\bibnamefont {Bamba}}, \bibinfo {author} {\bibfnamefont {P.}~\bibnamefont {Senellart}}, \bibinfo {author} {\bibfnamefont {I.}~\bibnamefont {Sagnes}}, \bibinfo {author} {\bibfnamefont {A.}~\bibnamefont {Lem{\`{a}}tre}}, \bibinfo {author} {\bibfnamefont {C.}~\bibnamefont {Ciuti}},\ and\ \bibinfo {author} {\bibfnamefont {J.}~\bibnamefont {Bloch}},\ }\bibfield  {title} {\bibinfo {title} {{Polariton parametric oscillation in a single micropillar cavity}},\ }\href {https://doi.org/10.1063/1.3466902} {\bibfield  {journal} {\bibinfo  {journal} {Appl. Phys. Lett.}\ }\textbf {\bibinfo {volume} {97}},\ \bibinfo {pages} {031105} (\bibinfo {year} {2010})}\BibitemShut {NoStop}%
\bibitem [{\citenamefont {Krizhanovskii}\ \emph {et~al.}(2013)\citenamefont {Krizhanovskii}, \citenamefont {Cerda-M{\'{e}}ndez}, \citenamefont {Gavrilov}, \citenamefont {Sarkar}, \citenamefont {Guda}, \citenamefont {Bradley}, \citenamefont {Santos}, \citenamefont {Hey}, \citenamefont {Biermann}, \citenamefont {Sich}, \citenamefont {Fras},\ and\ \citenamefont {Skolnick}}]{Krizhanovskii2013}%
  \BibitemOpen
  \bibfield  {author} {\bibinfo {author} {\bibfnamefont {D.~N.}\ \bibnamefont {Krizhanovskii}}, \bibinfo {author} {\bibfnamefont {E.~A.}\ \bibnamefont {Cerda-M{\'{e}}ndez}}, \bibinfo {author} {\bibfnamefont {S.}~\bibnamefont {Gavrilov}}, \bibinfo {author} {\bibfnamefont {D.}~\bibnamefont {Sarkar}}, \bibinfo {author} {\bibfnamefont {K.}~\bibnamefont {Guda}}, \bibinfo {author} {\bibfnamefont {R.}~\bibnamefont {Bradley}}, \bibinfo {author} {\bibfnamefont {P.~V.}\ \bibnamefont {Santos}}, \bibinfo {author} {\bibfnamefont {R.}~\bibnamefont {Hey}}, \bibinfo {author} {\bibfnamefont {K.}~\bibnamefont {Biermann}}, \bibinfo {author} {\bibfnamefont {M.}~\bibnamefont {Sich}}, \bibinfo {author} {\bibfnamefont {F.}~\bibnamefont {Fras}},\ and\ \bibinfo {author} {\bibfnamefont {M.~S.}\ \bibnamefont {Skolnick}},\ }\bibfield  {title} {\bibinfo {title} {{Effect of polariton-polariton interactions on the excitation spectrum of a nonequilibrium condensate in a periodic potential}},\ }\href
  {https://doi.org/10.1103/PhysRevB.87.155423} {\bibfield  {journal} {\bibinfo  {journal} {Phys. Rev. B}\ }\textbf {\bibinfo {volume} {87}},\ \bibinfo {pages} {155423} (\bibinfo {year} {2013})}\BibitemShut {NoStop}%
\bibitem [{\citenamefont {Kuznetsov}\ \emph {et~al.}(2020)\citenamefont {Kuznetsov}, \citenamefont {Dagvadorj}, \citenamefont {Biermann}, \citenamefont {Szymanska},\ and\ \citenamefont {Santos}}]{Kuznetsov2020b}%
  \BibitemOpen
  \bibfield  {author} {\bibinfo {author} {\bibfnamefont {A.~S.}\ \bibnamefont {Kuznetsov}}, \bibinfo {author} {\bibfnamefont {G.}~\bibnamefont {Dagvadorj}}, \bibinfo {author} {\bibfnamefont {K.}~\bibnamefont {Biermann}}, \bibinfo {author} {\bibfnamefont {M.}~\bibnamefont {Szymanska}},\ and\ \bibinfo {author} {\bibfnamefont {P.~V.}\ \bibnamefont {Santos}},\ }\bibfield  {title} {\bibinfo {title} {{Dynamically tuned arrays of polariton parametric oscillators}},\ }\href {https://doi.org/10.1364/optica.399747} {\bibfield  {journal} {\bibinfo  {journal} {Optica}\ }\textbf {\bibinfo {volume} {7}},\ \bibinfo {pages} {1673} (\bibinfo {year} {2020})}\BibitemShut {NoStop}%
\bibitem [{\citenamefont {Oka}\ and\ \citenamefont {Kitamura}(2019)}]{Oka2019}%
  \BibitemOpen
  \bibfield  {author} {\bibinfo {author} {\bibfnamefont {T.}~\bibnamefont {Oka}}\ and\ \bibinfo {author} {\bibfnamefont {S.}~\bibnamefont {Kitamura}},\ }\bibfield  {title} {\bibinfo {title} {{Floquet engineering of quantum materials}},\ }\href {https://doi.org/10.1146/annurev-conmatphys-031218-013423} {\bibfield  {journal} {\bibinfo  {journal} {Annu. Rev. Condens. Matter Phys.}\ }\textbf {\bibinfo {volume} {10}},\ \bibinfo {pages} {387} (\bibinfo {year} {2019})}\BibitemShut {NoStop}%
\bibitem [{\citenamefont {Silveri}\ \emph {et~al.}(2017)\citenamefont {Silveri}, \citenamefont {Tuorila}, \citenamefont {Thuneberg},\ and\ \citenamefont {Paraoanu}}]{Silveri2017}%
  \BibitemOpen
  \bibfield  {author} {\bibinfo {author} {\bibfnamefont {M.~P.}\ \bibnamefont {Silveri}}, \bibinfo {author} {\bibfnamefont {J.~A.}\ \bibnamefont {Tuorila}}, \bibinfo {author} {\bibfnamefont {E.~V.}\ \bibnamefont {Thuneberg}},\ and\ \bibinfo {author} {\bibfnamefont {G.~S.}\ \bibnamefont {Paraoanu}},\ }\bibfield  {title} {\bibinfo {title} {{Quantum systems under frequency modulation}},\ }\href {https://doi.org/10.1088/1361-6633/aa5170} {\bibfield  {journal} {\bibinfo  {journal} {Reports Prog. Phys.}\ }\textbf {\bibinfo {volume} {80}},\ \bibinfo {pages} {056002} (\bibinfo {year} {2017})}\BibitemShut {NoStop}%
\bibitem [{\citenamefont {Ivakhnenko}\ \emph {et~al.}(2023)\citenamefont {Ivakhnenko}, \citenamefont {Shevchenko},\ and\ \citenamefont {Nori}}]{Ivakhnenko2023}%
  \BibitemOpen
  \bibfield  {author} {\bibinfo {author} {\bibfnamefont {O.~V.}\ \bibnamefont {Ivakhnenko}}, \bibinfo {author} {\bibfnamefont {S.~N.}\ \bibnamefont {Shevchenko}},\ and\ \bibinfo {author} {\bibfnamefont {F.}~\bibnamefont {Nori}},\ }\bibfield  {title} {\bibinfo {title} {{Nonadiabatic Landau–Zener–St{\"{u}}ckelberg–Majorana transitions, dynamics, and interference}},\ }\href {https://doi.org/10.1016/j.physrep.2022.10.002} {\bibfield  {journal} {\bibinfo  {journal} {Phys. Rep.}\ }\textbf {\bibinfo {volume} {995}},\ \bibinfo {pages} {1} (\bibinfo {year} {2023})}\BibitemShut {NoStop}%
\bibitem [{\citenamefont {Erglis}\ \emph {et~al.}(2025)\citenamefont {Erglis}, \citenamefont {Sazhin}, \citenamefont {Vewinger}, \citenamefont {Weitz}, \citenamefont {Buhmann},\ and\ \citenamefont {Schmitt}}]{Erglis2025}%
  \BibitemOpen
  \bibfield  {author} {\bibinfo {author} {\bibfnamefont {A.}~\bibnamefont {Erglis}}, \bibinfo {author} {\bibfnamefont {A.}~\bibnamefont {Sazhin}}, \bibinfo {author} {\bibfnamefont {F.}~\bibnamefont {Vewinger}}, \bibinfo {author} {\bibfnamefont {M.}~\bibnamefont {Weitz}}, \bibinfo {author} {\bibfnamefont {S.~Y.}\ \bibnamefont {Buhmann}},\ and\ \bibinfo {author} {\bibfnamefont {J.}~\bibnamefont {Schmitt}},\ }\bibfield  {title} {\bibinfo {title} {{Time-periodic driving of a bath-coupled open quantum gas of light}},\ }\href@noop {} {\bibfield  {journal} {\bibinfo  {journal} {ArXiv}\ } (\bibinfo {year} {2025})},\ \Eprint {https://arxiv.org/abs/2502.14986} {arXiv:2502.14986} \BibitemShut {NoStop}%
\bibitem [{\citenamefont {{De Lima}}\ and\ \citenamefont {Santos}(2005)}]{DeLima2005}%
  \BibitemOpen
  \bibfield  {author} {\bibinfo {author} {\bibfnamefont {M.~M.}\ \bibnamefont {{De Lima}}}\ and\ \bibinfo {author} {\bibfnamefont {P.~V.}\ \bibnamefont {Santos}},\ }\bibfield  {title} {\bibinfo {title} {{Modulation of photonic structures by surface acoustic waves}},\ }\href {https://doi.org/10.1088/0034-4885/68/7/R02} {\bibfield  {journal} {\bibinfo  {journal} {Reports Prog. Phys.}\ }\textbf {\bibinfo {volume} {68}},\ \bibinfo {pages} {1639} (\bibinfo {year} {2005})}\BibitemShut {NoStop}%
\bibitem [{\citenamefont {Couto}\ \emph {et~al.}(2009)\citenamefont {Couto}, \citenamefont {Lazic}, \citenamefont {Iikawa}, \citenamefont {Stotz}, \citenamefont {Jahn}, \citenamefont {Hey},\ and\ \citenamefont {Santos}}]{Couto2009}%
  \BibitemOpen
  \bibfield  {author} {\bibinfo {author} {\bibfnamefont {O.~D.~D.}\ \bibnamefont {Couto}}, \bibinfo {author} {\bibfnamefont {S.}~\bibnamefont {Lazic}}, \bibinfo {author} {\bibfnamefont {F.}~\bibnamefont {Iikawa}}, \bibinfo {author} {\bibfnamefont {J.~A.~H.}\ \bibnamefont {Stotz}}, \bibinfo {author} {\bibfnamefont {U.}~\bibnamefont {Jahn}}, \bibinfo {author} {\bibfnamefont {R.}~\bibnamefont {Hey}},\ and\ \bibinfo {author} {\bibfnamefont {P.~V.}\ \bibnamefont {Santos}},\ }\bibfield  {title} {\bibinfo {title} {{Photon anti-bunching in acoustically pumped quantum dots}},\ }\href {https://doi.org/10.1038/NPHOTON.2009.191} {\bibfield  {journal} {\bibinfo  {journal} {Nat. Photonics}\ }\textbf {\bibinfo {volume} {3}},\ \bibinfo {pages} {645} (\bibinfo {year} {2009})}\BibitemShut {NoStop}%
\bibitem [{\citenamefont {Berstermann}\ \emph {et~al.}(2009)\citenamefont {Berstermann}, \citenamefont {Scherbakov}, \citenamefont {Akimov}, \citenamefont {Yakovlev}, \citenamefont {Gippius}, \citenamefont {Glavin}, \citenamefont {Sagnes}, \citenamefont {Bloch},\ and\ \citenamefont {Bayer}}]{Berstermann2009}%
  \BibitemOpen
  \bibfield  {author} {\bibinfo {author} {\bibfnamefont {T.}~\bibnamefont {Berstermann}}, \bibinfo {author} {\bibfnamefont {A.~V.}\ \bibnamefont {Scherbakov}}, \bibinfo {author} {\bibfnamefont {A.~V.}\ \bibnamefont {Akimov}}, \bibinfo {author} {\bibfnamefont {D.~R.}\ \bibnamefont {Yakovlev}}, \bibinfo {author} {\bibfnamefont {N.~A.}\ \bibnamefont {Gippius}}, \bibinfo {author} {\bibfnamefont {B.~A.}\ \bibnamefont {Glavin}}, \bibinfo {author} {\bibfnamefont {I.}~\bibnamefont {Sagnes}}, \bibinfo {author} {\bibfnamefont {J.}~\bibnamefont {Bloch}},\ and\ \bibinfo {author} {\bibfnamefont {M.}~\bibnamefont {Bayer}},\ }\bibfield  {title} {\bibinfo {title} {{Terahertz polariton sidebands generated by ultrafast strain pulses in an optical semiconductor microcavity}},\ }\href {https://doi.org/10.1103/PhysRevB.80.075301} {\bibfield  {journal} {\bibinfo  {journal} {Phys. Rev. B}\ }\textbf {\bibinfo {volume} {80}},\ \bibinfo {pages} {075301} (\bibinfo {year} {2009})}\BibitemShut {NoStop}%
\bibitem [{\citenamefont {Fuhrmann}\ \emph {et~al.}(2011)\citenamefont {Fuhrmann}, \citenamefont {Thon}, \citenamefont {Kim}, \citenamefont {Bouwmeester}, \citenamefont {Petroff}, \citenamefont {Wixforth},\ and\ \citenamefont {Krenner}}]{Fuhrmann2011}%
  \BibitemOpen
  \bibfield  {author} {\bibinfo {author} {\bibfnamefont {D.~A.}\ \bibnamefont {Fuhrmann}}, \bibinfo {author} {\bibfnamefont {S.~M.}\ \bibnamefont {Thon}}, \bibinfo {author} {\bibfnamefont {H.}~\bibnamefont {Kim}}, \bibinfo {author} {\bibfnamefont {D.}~\bibnamefont {Bouwmeester}}, \bibinfo {author} {\bibfnamefont {P.~M.}\ \bibnamefont {Petroff}}, \bibinfo {author} {\bibfnamefont {A.}~\bibnamefont {Wixforth}},\ and\ \bibinfo {author} {\bibfnamefont {H.~J.}\ \bibnamefont {Krenner}},\ }\bibfield  {title} {\bibinfo {title} {{Dynamic modulation of photonic crystal nanocavities using gigahertz acoustic phonons}},\ }\href {https://doi.org/10.1038/nphoton.2011.208} {\bibfield  {journal} {\bibinfo  {journal} {Nat. Photonics}\ }\textbf {\bibinfo {volume} {5}},\ \bibinfo {pages} {605} (\bibinfo {year} {2011})}\BibitemShut {NoStop}%
\bibitem [{\citenamefont {Jusserand}\ \emph {et~al.}(2015)\citenamefont {Jusserand}, \citenamefont {Poddubny}, \citenamefont {Poshakinskiy}, \citenamefont {Fainstein},\ and\ \citenamefont {Lemaitre}}]{Jusserand2015}%
  \BibitemOpen
  \bibfield  {author} {\bibinfo {author} {\bibfnamefont {B.}~\bibnamefont {Jusserand}}, \bibinfo {author} {\bibfnamefont {A.~N.}\ \bibnamefont {Poddubny}}, \bibinfo {author} {\bibfnamefont {A.~V.}\ \bibnamefont {Poshakinskiy}}, \bibinfo {author} {\bibfnamefont {A.}~\bibnamefont {Fainstein}},\ and\ \bibinfo {author} {\bibfnamefont {A.}~\bibnamefont {Lemaitre}},\ }\bibfield  {title} {\bibinfo {title} {{Polariton Resonances for Ultrastrong Coupling Cavity Optomechanics in GaAs/AlAs Multiple Quantum Wells}},\ }\href {https://doi.org/10.1103/PhysRevLett.115.267402} {\bibfield  {journal} {\bibinfo  {journal} {Phys. Rev. Lett.}\ }\textbf {\bibinfo {volume} {115}},\ \bibinfo {pages} {267402} (\bibinfo {year} {2015})}\BibitemShut {NoStop}%
\bibitem [{\citenamefont {Wei{\ss}}\ \emph {et~al.}(2016)\citenamefont {Wei{\ss}}, \citenamefont {Kapfinger}, \citenamefont {Reichert}, \citenamefont {Finley}, \citenamefont {Wixforth}, \citenamefont {Kaniber},\ and\ \citenamefont {Krenner}}]{Weiss2016}%
  \BibitemOpen
  \bibfield  {author} {\bibinfo {author} {\bibfnamefont {M.}~\bibnamefont {Wei{\ss}}}, \bibinfo {author} {\bibfnamefont {S.}~\bibnamefont {Kapfinger}}, \bibinfo {author} {\bibfnamefont {T.}~\bibnamefont {Reichert}}, \bibinfo {author} {\bibfnamefont {J.~J.}\ \bibnamefont {Finley}}, \bibinfo {author} {\bibfnamefont {A.}~\bibnamefont {Wixforth}}, \bibinfo {author} {\bibfnamefont {M.}~\bibnamefont {Kaniber}},\ and\ \bibinfo {author} {\bibfnamefont {H.~J.}\ \bibnamefont {Krenner}},\ }\bibfield  {title} {\bibinfo {title} {{Surface acoustic wave regulated single photon emission from a coupled quantum dot–nanocavity system}},\ }\href {https://doi.org/10.1063/1.4959079} {\bibfield  {journal} {\bibinfo  {journal} {Appl. Phys. Lett.}\ }\textbf {\bibinfo {volume} {109}},\ \bibinfo {pages} {033105} (\bibinfo {year} {2016})}\BibitemShut {NoStop}%
\bibitem [{\citenamefont {Hern{\'{a}}ndez-M{\'{i}}nguez}\ \emph {et~al.}(2020)\citenamefont {Hern{\'{a}}ndez-M{\'{i}}nguez}, \citenamefont {Poshakinskiy}, \citenamefont {Hollenbach}, \citenamefont {Santos},\ and\ \citenamefont {Astakhov}}]{Hernandez-Minguez2020}%
  \BibitemOpen
  \bibfield  {author} {\bibinfo {author} {\bibfnamefont {A.}~\bibnamefont {Hern{\'{a}}ndez-M{\'{i}}nguez}}, \bibinfo {author} {\bibfnamefont {A.~V.}\ \bibnamefont {Poshakinskiy}}, \bibinfo {author} {\bibfnamefont {M.}~\bibnamefont {Hollenbach}}, \bibinfo {author} {\bibfnamefont {P.~V.}\ \bibnamefont {Santos}},\ and\ \bibinfo {author} {\bibfnamefont {G.~V.}\ \bibnamefont {Astakhov}},\ }\bibfield  {title} {\bibinfo {title} {{Anisotropic Spin-Acoustic Resonance in Silicon Carbide at Room Temperature}},\ }\href {https://doi.org/10.1103/PhysRevLett.125.107702} {\bibfield  {journal} {\bibinfo  {journal} {Phys. Rev. Lett.}\ }\textbf {\bibinfo {volume} {125}},\ \bibinfo {pages} {107702} (\bibinfo {year} {2020})}\BibitemShut {NoStop}%
\bibitem [{\citenamefont {Sesin}\ \emph {et~al.}(2023)\citenamefont {Sesin}, \citenamefont {Kuznetsov}, \citenamefont {Rozas}, \citenamefont {Anguiano}, \citenamefont {Bruchhausen}, \citenamefont {Lema{\^{i}}tre}, \citenamefont {Biermann}, \citenamefont {Santos},\ and\ \citenamefont {Fainstein}}]{Sesin2023}%
  \BibitemOpen
  \bibfield  {author} {\bibinfo {author} {\bibfnamefont {P.}~\bibnamefont {Sesin}}, \bibinfo {author} {\bibfnamefont {A.~S.}\ \bibnamefont {Kuznetsov}}, \bibinfo {author} {\bibfnamefont {G.}~\bibnamefont {Rozas}}, \bibinfo {author} {\bibfnamefont {S.}~\bibnamefont {Anguiano}}, \bibinfo {author} {\bibfnamefont {A.~E.}\ \bibnamefont {Bruchhausen}}, \bibinfo {author} {\bibfnamefont {A.}~\bibnamefont {Lema{\^{i}}tre}}, \bibinfo {author} {\bibfnamefont {K.}~\bibnamefont {Biermann}}, \bibinfo {author} {\bibfnamefont {P.~V.}\ \bibnamefont {Santos}},\ and\ \bibinfo {author} {\bibfnamefont {A.}~\bibnamefont {Fainstein}},\ }\bibfield  {title} {\bibinfo {title} {{Giant optomechanical coupling and dephasing protection with cavity exciton-polaritons}},\ }\href {https://doi.org/10.1103/PhysRevResearch.5.L042035} {\bibfield  {journal} {\bibinfo  {journal} {Phys. Rev. Res.}\ }\textbf {\bibinfo {volume} {5}},\ \bibinfo {pages} {L042035} (\bibinfo {year} {2023})}\BibitemShut {NoStop}%
\bibitem [{\citenamefont {Berstermann}\ \emph {et~al.}(2012)\citenamefont {Berstermann}, \citenamefont {Br{\"{u}}ggemann}, \citenamefont {Akimov}, \citenamefont {Bombeck}, \citenamefont {Yakovlev}, \citenamefont {Gippius}, \citenamefont {Scherbakov}, \citenamefont {Sagnes}, \citenamefont {Bloch},\ and\ \citenamefont {Bayer}}]{Berstermann2012}%
  \BibitemOpen
  \bibfield  {author} {\bibinfo {author} {\bibfnamefont {T.}~\bibnamefont {Berstermann}}, \bibinfo {author} {\bibfnamefont {C.}~\bibnamefont {Br{\"{u}}ggemann}}, \bibinfo {author} {\bibfnamefont {A.~V.}\ \bibnamefont {Akimov}}, \bibinfo {author} {\bibfnamefont {M.}~\bibnamefont {Bombeck}}, \bibinfo {author} {\bibfnamefont {D.~R.}\ \bibnamefont {Yakovlev}}, \bibinfo {author} {\bibfnamefont {N.~A.}\ \bibnamefont {Gippius}}, \bibinfo {author} {\bibfnamefont {A.~V.}\ \bibnamefont {Scherbakov}}, \bibinfo {author} {\bibfnamefont {I.}~\bibnamefont {Sagnes}}, \bibinfo {author} {\bibfnamefont {J.}~\bibnamefont {Bloch}},\ and\ \bibinfo {author} {\bibfnamefont {M.}~\bibnamefont {Bayer}},\ }\bibfield  {title} {\bibinfo {title} {{Destruction and recurrence of excitons by acoustic shock waves on picosecond time scales}},\ }\href {https://doi.org/10.1103/PhysRevB.86.195306} {\bibfield  {journal} {\bibinfo  {journal} {Phys. Rev. B}\ }\textbf {\bibinfo {volume} {86}},\ \bibinfo {pages} {195306} (\bibinfo {year}
  {2012})}\BibitemShut {NoStop}%
\bibitem [{\citenamefont {Fainstein}\ \emph {et~al.}(2013)\citenamefont {Fainstein}, \citenamefont {Lanzillotti-Kimura}, \citenamefont {Jusserand},\ and\ \citenamefont {Perrin}}]{Fainstein2013}%
  \BibitemOpen
  \bibfield  {author} {\bibinfo {author} {\bibfnamefont {A.}~\bibnamefont {Fainstein}}, \bibinfo {author} {\bibfnamefont {N.~D.}\ \bibnamefont {Lanzillotti-Kimura}}, \bibinfo {author} {\bibfnamefont {B.}~\bibnamefont {Jusserand}},\ and\ \bibinfo {author} {\bibfnamefont {B.}~\bibnamefont {Perrin}},\ }\bibfield  {title} {\bibinfo {title} {{Strong optical-mechanical coupling in a vertical GaAs/AlAs microcavity for subterahertz phonons and near-infrared light}},\ }\href {https://doi.org/10.1103/PhysRevLett.110.037403} {\bibfield  {journal} {\bibinfo  {journal} {Phys. Rev. Lett.}\ }\textbf {\bibinfo {volume} {110}},\ \bibinfo {pages} {037403} (\bibinfo {year} {2013})}\BibitemShut {NoStop}%
\bibitem [{\citenamefont {{El Daif}}\ \emph {et~al.}(2006)\citenamefont {{El Daif}}, \citenamefont {Baas}, \citenamefont {Guillet}, \citenamefont {Brantut}, \citenamefont {Kaitouni}, \citenamefont {Staehli}, \citenamefont {Morier-Genoud},\ and\ \citenamefont {Deveaud}}]{ElDaif2006}%
  \BibitemOpen
  \bibfield  {author} {\bibinfo {author} {\bibfnamefont {O.}~\bibnamefont {{El Daif}}}, \bibinfo {author} {\bibfnamefont {A.}~\bibnamefont {Baas}}, \bibinfo {author} {\bibfnamefont {T.}~\bibnamefont {Guillet}}, \bibinfo {author} {\bibfnamefont {J.-P.}\ \bibnamefont {Brantut}}, \bibinfo {author} {\bibfnamefont {R.~I.}\ \bibnamefont {Kaitouni}}, \bibinfo {author} {\bibfnamefont {J.~L.}\ \bibnamefont {Staehli}}, \bibinfo {author} {\bibfnamefont {F.}~\bibnamefont {Morier-Genoud}},\ and\ \bibinfo {author} {\bibfnamefont {B.}~\bibnamefont {Deveaud}},\ }\bibfield  {title} {\bibinfo {title} {{Polariton quantum boxes in semiconductor microcavities}},\ }\href {https://doi.org/10.1063/1.2172409} {\bibfield  {journal} {\bibinfo  {journal} {Appl. Phys. Lett.}\ }\textbf {\bibinfo {volume} {88}},\ \bibinfo {pages} {061105} (\bibinfo {year} {2006})}\BibitemShut {NoStop}%
\bibitem [{\citenamefont {Machado}\ \emph {et~al.}(2019)\citenamefont {Machado}, \citenamefont {Crespo-Poveda}, \citenamefont {Kuznetsov}, \citenamefont {Biermann}, \citenamefont {Scalvi},\ and\ \citenamefont {Santos}}]{Machado2019}%
  \BibitemOpen
  \bibfield  {author} {\bibinfo {author} {\bibfnamefont {D.~H.~O.}\ \bibnamefont {Machado}}, \bibinfo {author} {\bibfnamefont {A.}~\bibnamefont {Crespo-Poveda}}, \bibinfo {author} {\bibfnamefont {A.~S.}\ \bibnamefont {Kuznetsov}}, \bibinfo {author} {\bibfnamefont {K.}~\bibnamefont {Biermann}}, \bibinfo {author} {\bibfnamefont {L.~V.~A.}\ \bibnamefont {Scalvi}},\ and\ \bibinfo {author} {\bibfnamefont {P.~V.}\ \bibnamefont {Santos}},\ }\bibfield  {title} {\bibinfo {title} {{Generation and propagation of super-high-frequency bulk acoustic wave in GaAs}},\ }\href {https://doi.org/10.1103/PhysRevApplied.12.044013} {\bibfield  {journal} {\bibinfo  {journal} {Phys. Rev. Appl.}\ }\textbf {\bibinfo {volume} {12}},\ \bibinfo {pages} {1} (\bibinfo {year} {2019})}\BibitemShut {NoStop}%
\bibitem [{\citenamefont {de~Lima}\ \emph {et~al.}(2006)\citenamefont {de~Lima}, \citenamefont {van~der Poel}, \citenamefont {Santos},\ and\ \citenamefont {Hvam}}]{DeLima2006}%
  \BibitemOpen
  \bibfield  {author} {\bibinfo {author} {\bibfnamefont {M.~M.}\ \bibnamefont {de~Lima}}, \bibinfo {author} {\bibfnamefont {M.}~\bibnamefont {van~der Poel}}, \bibinfo {author} {\bibfnamefont {P.~V.}\ \bibnamefont {Santos}},\ and\ \bibinfo {author} {\bibfnamefont {J.~M.}\ \bibnamefont {Hvam}},\ }\bibfield  {title} {\bibinfo {title} {{Phonon-induced polariton superlattices}},\ }\href {https://doi.org/10.1103/PhysRevLett.97.045501} {\bibfield  {journal} {\bibinfo  {journal} {Phys. Rev. Lett.}\ }\textbf {\bibinfo {volume} {97}},\ \bibinfo {pages} {045501} (\bibinfo {year} {2006})}\BibitemShut {NoStop}%
\bibitem [{\citenamefont {Cerda-M{\'{e}}ndez}\ \emph {et~al.}(2010)\citenamefont {Cerda-M{\'{e}}ndez}, \citenamefont {Krizhanovskii}, \citenamefont {Wouters}, \citenamefont {Bradley}, \citenamefont {Biermann}, \citenamefont {Guda}, \citenamefont {Hey}, \citenamefont {Santos}, \citenamefont {Sarkar},\ and\ \citenamefont {Skolnick}}]{Cerda-Mendez2010}%
  \BibitemOpen
  \bibfield  {author} {\bibinfo {author} {\bibfnamefont {E.~A.}\ \bibnamefont {Cerda-M{\'{e}}ndez}}, \bibinfo {author} {\bibfnamefont {D.~N.}\ \bibnamefont {Krizhanovskii}}, \bibinfo {author} {\bibfnamefont {M.}~\bibnamefont {Wouters}}, \bibinfo {author} {\bibfnamefont {R.}~\bibnamefont {Bradley}}, \bibinfo {author} {\bibfnamefont {K.}~\bibnamefont {Biermann}}, \bibinfo {author} {\bibfnamefont {K.}~\bibnamefont {Guda}}, \bibinfo {author} {\bibfnamefont {R.}~\bibnamefont {Hey}}, \bibinfo {author} {\bibfnamefont {P.~V.}\ \bibnamefont {Santos}}, \bibinfo {author} {\bibfnamefont {D.}~\bibnamefont {Sarkar}},\ and\ \bibinfo {author} {\bibfnamefont {M.~S.}\ \bibnamefont {Skolnick}},\ }\bibfield  {title} {\bibinfo {title} {{Polariton condensation in dynamic acoustic lattices}},\ }\href {https://doi.org/10.1103/PhysRevLett.105.116402} {\bibfield  {journal} {\bibinfo  {journal} {Phys. Rev. Lett.}\ }\textbf {\bibinfo {volume} {105}},\ \bibinfo {pages} {116402} (\bibinfo {year} {2010})}\BibitemShut {NoStop}%
\bibitem [{\citenamefont {Kuznetsov}\ \emph {et~al.}(2021)\citenamefont {Kuznetsov}, \citenamefont {Machado}, \citenamefont {Biermann},\ and\ \citenamefont {Santos}}]{Kuznetsov2021}%
  \BibitemOpen
  \bibfield  {author} {\bibinfo {author} {\bibfnamefont {A.~S.}\ \bibnamefont {Kuznetsov}}, \bibinfo {author} {\bibfnamefont {D.~H.~O.}\ \bibnamefont {Machado}}, \bibinfo {author} {\bibfnamefont {K.}~\bibnamefont {Biermann}},\ and\ \bibinfo {author} {\bibfnamefont {P.~V.}\ \bibnamefont {Santos}},\ }\bibfield  {title} {\bibinfo {title} {{Electrically Driven Microcavity Exciton-Polariton Optomechanics at 20 GHz}},\ }\href {https://doi.org/10.1103/PhysRevX.11.021020} {\bibfield  {journal} {\bibinfo  {journal} {Phys. Rev. X}\ }\textbf {\bibinfo {volume} {11}},\ \bibinfo {pages} {021020} (\bibinfo {year} {2021})}\BibitemShut {NoStop}%
\bibitem [{\citenamefont {Kuznetsov}\ \emph {et~al.}(2018)\citenamefont {Kuznetsov}, \citenamefont {Helgers}, \citenamefont {Biermann},\ and\ \citenamefont {Santos}}]{Kuznetsov2018}%
  \BibitemOpen
  \bibfield  {author} {\bibinfo {author} {\bibfnamefont {A.~S.}\ \bibnamefont {Kuznetsov}}, \bibinfo {author} {\bibfnamefont {P.~L.~J.}\ \bibnamefont {Helgers}}, \bibinfo {author} {\bibfnamefont {K.}~\bibnamefont {Biermann}},\ and\ \bibinfo {author} {\bibfnamefont {P.~V.}\ \bibnamefont {Santos}},\ }\bibfield  {title} {\bibinfo {title} {{Quantum confinement of exciton-polaritons in a structured (Al,Ga)As microcavity}},\ }\href {https://doi.org/10.1103/PhysRevB.97.195309} {\bibfield  {journal} {\bibinfo  {journal} {Phys. Rev. B}\ }\textbf {\bibinfo {volume} {97}},\ \bibinfo {pages} {195309} (\bibinfo {year} {2018})}\BibitemShut {NoStop}%
\bibitem [{\citenamefont {Kuznetsov}\ \emph {et~al.}(2023)\citenamefont {Kuznetsov}, \citenamefont {Biermann}, \citenamefont {Reynoso}, \citenamefont {Fainstein},\ and\ \citenamefont {Santos}}]{Kuznetsov2023}%
  \BibitemOpen
  \bibfield  {author} {\bibinfo {author} {\bibfnamefont {A.~S.}\ \bibnamefont {Kuznetsov}}, \bibinfo {author} {\bibfnamefont {K.}~\bibnamefont {Biermann}}, \bibinfo {author} {\bibfnamefont {A.~A.}\ \bibnamefont {Reynoso}}, \bibinfo {author} {\bibfnamefont {A.}~\bibnamefont {Fainstein}},\ and\ \bibinfo {author} {\bibfnamefont {P.~V.}\ \bibnamefont {Santos}},\ }\bibfield  {title} {\bibinfo {title} {{Microcavity phonoritons – a coherent optical-to-microwave interface}},\ }\href {https://doi.org/10.1038/s41467-023-40894-7} {\bibfield  {journal} {\bibinfo  {journal} {Nat. Commun.}\ }\textbf {\bibinfo {volume} {14}},\ \bibinfo {pages} {5470} (\bibinfo {year} {2023})}\BibitemShut {NoStop}%
\bibitem [{\citenamefont {Trigo}\ \emph {et~al.}(2002)\citenamefont {Trigo}, \citenamefont {Bruchhausen}, \citenamefont {Fainstein}, \citenamefont {Jusserand},\ and\ \citenamefont {Thierry-Mieg}}]{Trigo2002}%
  \BibitemOpen
  \bibfield  {author} {\bibinfo {author} {\bibfnamefont {M.}~\bibnamefont {Trigo}}, \bibinfo {author} {\bibfnamefont {A.}~\bibnamefont {Bruchhausen}}, \bibinfo {author} {\bibfnamefont {A.}~\bibnamefont {Fainstein}}, \bibinfo {author} {\bibfnamefont {B.}~\bibnamefont {Jusserand}},\ and\ \bibinfo {author} {\bibfnamefont {V.}~\bibnamefont {Thierry-Mieg}},\ }\bibfield  {title} {\bibinfo {title} {{Confinement of acoustical vibrations in a semiconductor planar phonon cavity}},\ }\href {https://doi.org/10.1103/PhysRevLett.89.227402} {\bibfield  {journal} {\bibinfo  {journal} {Phys. Rev. Lett.}\ }\textbf {\bibinfo {volume} {89}},\ \bibinfo {pages} {227402} (\bibinfo {year} {2002})}\BibitemShut {NoStop}%
\end{thebibliography}%


\begin{thebibliography}{6}%
\makeatletter
\providecommand \@ifxundefined [1]{%
 \@ifx{#1\undefined}
}%
\providecommand \@ifnum [1]{%
 \ifnum #1\expandafter \@firstoftwo
 \else \expandafter \@secondoftwo
 \fi
}%
\providecommand \@ifx [1]{%
 \ifx #1\expandafter \@firstoftwo
 \else \expandafter \@secondoftwo
 \fi
}%
\providecommand \natexlab [1]{#1}%
\providecommand \enquote  [1]{``#1''}%
\providecommand \bibnamefont  [1]{#1}%
\providecommand \bibfnamefont [1]{#1}%
\providecommand \citenamefont [1]{#1}%
\providecommand \href@noop [0]{\@secondoftwo}%
\providecommand \href [0]{\begingroup \@sanitize@url \@href}%
\providecommand \@href[1]{\@@startlink{#1}\@@href}%
\providecommand \@@href[1]{\endgroup#1\@@endlink}%
\providecommand \@sanitize@url [0]{\catcode `\\12\catcode `\$12\catcode `\&12\catcode `\#12\catcode `\^12\catcode `\_12\catcode `\%12\relax}%
\providecommand \@@startlink[1]{}%
\providecommand \@@endlink[0]{}%
\providecommand \url  [0]{\begingroup\@sanitize@url \@url }%
\providecommand \@url [1]{\endgroup\@href {#1}{\urlprefix }}%
\providecommand \urlprefix  [0]{URL }%
\providecommand \Eprint [0]{\href }%
\providecommand \doibase [0]{https://doi.org/}%
\providecommand \selectlanguage [0]{\@gobble}%
\providecommand \bibinfo  [0]{\@secondoftwo}%
\providecommand \bibfield  [0]{\@secondoftwo}%
\providecommand \translation [1]{[#1]}%
\providecommand \BibitemOpen [0]{}%
\providecommand \bibitemStop [0]{}%
\providecommand \bibitemNoStop [0]{.\EOS\space}%
\providecommand \EOS [0]{\spacefactor3000\relax}%
\providecommand \BibitemShut  [1]{\csname bibitem#1\endcsname}%
\let\auto@bib@innerbib\@empty
\bibitem [{\citenamefont {Kuznetsov}\ \emph {et~al.}(2023)\citenamefont {Kuznetsov}, \citenamefont {Biermann}, \citenamefont {Reynoso}, \citenamefont {Fainstein},\ and\ \citenamefont {Santos}}]{Kuznetsov2023}%
  \BibitemOpen
  \bibfield  {author} {\bibinfo {author} {\bibfnamefont {A.~S.}\ \bibnamefont {Kuznetsov}}, \bibinfo {author} {\bibfnamefont {K.}~\bibnamefont {Biermann}}, \bibinfo {author} {\bibfnamefont {A.~A.}\ \bibnamefont {Reynoso}}, \bibinfo {author} {\bibfnamefont {A.}~\bibnamefont {Fainstein}},\ and\ \bibinfo {author} {\bibfnamefont {P.~V.}\ \bibnamefont {Santos}},\ }\bibfield  {title} {\bibinfo {title} {{Microcavity phonoritons – a coherent optical-to-microwave interface}},\ }\href {https://doi.org/10.1038/s41467-023-40894-7} {\bibfield  {journal} {\bibinfo  {journal} {Nat. Commun.}\ }\textbf {\bibinfo {volume} {14}},\ \bibinfo {pages} {5470} (\bibinfo {year} {2023})}\BibitemShut {NoStop}%
\bibitem [{\citenamefont {Eastham}(2008)}]{Eastham2008}%
  \BibitemOpen
  \bibfield  {author} {\bibinfo {author} {\bibfnamefont {P.~R.}\ \bibnamefont {Eastham}},\ }\bibfield  {title} {\bibinfo {title} {{Mode-locking and mode-competition in a non-equilibrium solid-state condensate}},\ }\href {https://doi.org/10.1103/PhysRevB.78.035319} {\bibfield  {journal} {\bibinfo  {journal} {Phys. Rev. B}\ }\textbf {\bibinfo {volume} {78}},\ \bibinfo {pages} {035319} (\bibinfo {year} {2008})}\BibitemShut {NoStop}%
\bibitem [{\citenamefont {Grosso}\ \emph {et~al.}(2014)\citenamefont {Grosso}, \citenamefont {Trebaol}, \citenamefont {Wouters}, \citenamefont {Morier-Genoud}, \citenamefont {Portella-Oberli},\ and\ \citenamefont {Deveaud}}]{Grosso2014}%
  \BibitemOpen
  \bibfield  {author} {\bibinfo {author} {\bibfnamefont {G.}~\bibnamefont {Grosso}}, \bibinfo {author} {\bibfnamefont {S.}~\bibnamefont {Trebaol}}, \bibinfo {author} {\bibfnamefont {M.}~\bibnamefont {Wouters}}, \bibinfo {author} {\bibfnamefont {F.}~\bibnamefont {Morier-Genoud}}, \bibinfo {author} {\bibfnamefont {M.~T.}\ \bibnamefont {Portella-Oberli}},\ and\ \bibinfo {author} {\bibfnamefont {B.}~\bibnamefont {Deveaud}},\ }\bibfield  {title} {\bibinfo {title} {{Nonlinear relaxation and selective polychromatic lasing of confined polaritons}},\ }\href {https://doi.org/10.1103/PhysRevB.90.045307} {\bibfield  {journal} {\bibinfo  {journal} {Phys. Rev. B - Condens. Matter Mater. Phys.}\ }\textbf {\bibinfo {volume} {90}},\ \bibinfo {pages} {045307} (\bibinfo {year} {2014})}\BibitemShut {NoStop}%
\bibitem [{\citenamefont {Kuznetsov}\ \emph {et~al.}(2024)\citenamefont {Kuznetsov}, \citenamefont {Biermann},\ and\ \citenamefont {Santos}}]{Kuznetsov2024}%
  \BibitemOpen
  \bibfield  {author} {\bibinfo {author} {\bibfnamefont {A.~S.}\ \bibnamefont {Kuznetsov}}, \bibinfo {author} {\bibfnamefont {K.}~\bibnamefont {Biermann}},\ and\ \bibinfo {author} {\bibfnamefont {P.~V.}\ \bibnamefont {Santos}},\ }\bibfield  {title} {\bibinfo {title} {{Acceleration-induced spectral beats in strongly driven harmonic oscillators}},\ }\href {https://doi.org/10.1038/s41467-024-49610-5} {\bibfield  {journal} {\bibinfo  {journal} {Nat. Commun.}\ }\textbf {\bibinfo {volume} {15}},\ \bibinfo {pages} {5343} (\bibinfo {year} {2024})}\BibitemShut {NoStop}%
\bibitem [{\citenamefont {Wouters}\ and\ \citenamefont {Carusotto}(2007)}]{Wouters2007}%
  \BibitemOpen
  \bibfield  {author} {\bibinfo {author} {\bibfnamefont {M.}~\bibnamefont {Wouters}}\ and\ \bibinfo {author} {\bibfnamefont {I.}~\bibnamefont {Carusotto}},\ }\bibfield  {title} {\bibinfo {title} {{Excitations in a nonequilibrium bose-einstein condensate of exciton polaritons}},\ }\href {https://doi.org/10.1103/PhysRevLett.99.140402} {\bibfield  {journal} {\bibinfo  {journal} {Phys. Rev. Lett.}\ }\textbf {\bibinfo {volume} {99}},\ \bibinfo {pages} {140402} (\bibinfo {year} {2007})}\BibitemShut {NoStop}%
\bibitem [{\citenamefont {Wouters}\ \emph {et~al.}(2010)\citenamefont {Wouters}, \citenamefont {Liew},\ and\ \citenamefont {Savona}}]{Wouters2010}%
  \BibitemOpen
  \bibfield  {author} {\bibinfo {author} {\bibfnamefont {M.}~\bibnamefont {Wouters}}, \bibinfo {author} {\bibfnamefont {T.~C.}\ \bibnamefont {Liew}},\ and\ \bibinfo {author} {\bibfnamefont {V.}~\bibnamefont {Savona}},\ }\bibfield  {title} {\bibinfo {title} {{Energy relaxation in one-dimensional polariton condensates}},\ }\href {https://doi.org/10.1103/PhysRevB.82.245315} {\bibfield  {journal} {\bibinfo  {journal} {Phys. Rev. B - Condens. Matter Mater. Phys.}\ }\textbf {\bibinfo {volume} {82}},\ \bibinfo {pages} {245315} (\bibinfo {year} {2010})}\BibitemShut {NoStop}%
\end{thebibliography}%


\section*{Acknowledgments} 
We acknowledge the funding from German DFG grant 359162958. 
The authors thank Valentino Pistore for a critical review of the manuscript. We acknowledge the technical support by R. Baumann, S. Rauwerdink and W. Anders. 

\section*{Author contributions}
A.S.K. and I.C.-H. performed the spectroscopic experiments and carried out the respective analysis using the model developed by I.C.-H., G.U. and A.F.  K.B. performed the MBE growth of the samples.  A.S.K. and P.V.S. have conceived the idea. A.S.K has written the manuscript with the input from other authors.

\section*{Competing interests}
The authors declare no competing interests.


\end{document}


\title{Supplementary Information:\\
Ground state exciton-polariton condensation by coherent Floquet driving}

\author{A. S. Kuznetsov}
\email[corresponding author: ]{kuznetsov@pdi-berlin.de}
\affiliation{Paul-Drude-Institut f{\"u}r Festk{\"o}rperelektronik, Leibniz-Institut im Forschungsverbund Berlin e. V., Hausvogteiplatz 5-7, 10117 Berlin, Germany}

\author{I. Carraro-Haddad}
\affiliation{Bariloche Atomic Centre and Balseiro Institute, National Council for Scientific and Technical Research, 8400 S.C. de Bariloche, R.N., Argentina}
\affiliation{Institute of Nanoscience and Nanotechnology, National Council for Scientific and Technical Research,8400 Bariloche, Argentina}

\author{G. Usaj}
\affiliation{Bariloche Atomic Centre and Balseiro Institute, National Council for Scientific and Technical Research, 8400 S.C. de Bariloche, R.N., Argentina}
\affiliation{Institute of Nanoscience and Nanotechnology, National Council for Scientific and Technical Research,8400 Bariloche, Argentina}

\author{K. Biermann}
\affiliation{Paul-Drude-Institut f{\"u}r Festk{\"o}rperelektronik, Leibniz-Institut im Forschungsverbund Berlin e. V., Hausvogteiplatz 5-7, 10117 Berlin, Germany}

\author{A. Fainstein}
\affiliation{Bariloche Atomic Centre and Balseiro Institute, National Council for Scientific and Technical Research, 8400 S.C. de Bariloche, R.N., Argentina}
\affiliation{Institute of Nanoscience and Nanotechnology, National Council for Scientific and Technical Research,8400 Bariloche, Argentina}

\author{P. V. Santos}
\affiliation{Paul-Drude-Institut f{\"u}r Festk{\"o}rperelektronik, Leibniz-Institut im Forschungsverbund Berlin e. V., Hausvogteiplatz 5-7, 10117 Berlin, Germany}

\maketitle



\vskip 1 cm
\centering{This document contains additional material to support the conclusions of the main text.}
\vskip 1 cm

\newpage
\section{Spatial dispersion of polariton energies}
\label{appendix:spatialDispersion}

\begin{figure*}[tbhp]
	\centering
	\includegraphics[width=0.7\textwidth, keepaspectratio=true]{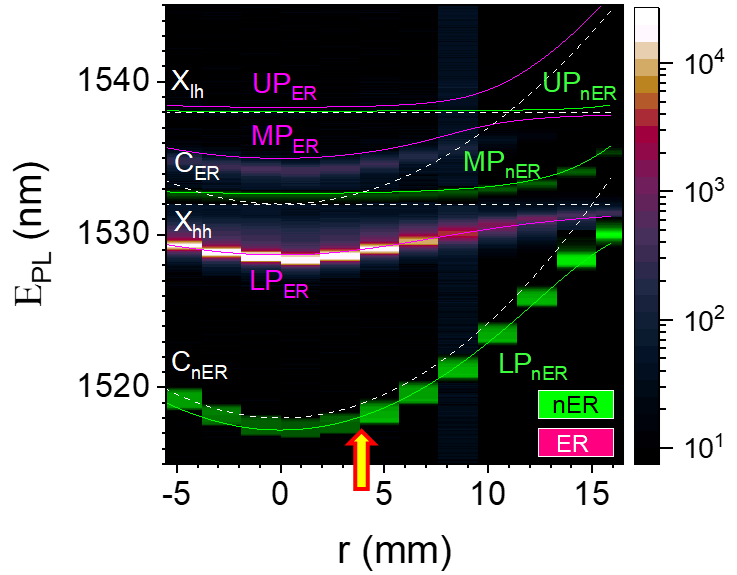}
	\caption{
        Spatial dispersion of polariton emission along the radial direction ($r$) of a 2" wafer for the non-etched (nER, green) and etched (ER, pink) regions. r = 0 corresponds to the wafer center. The solid lines are the fits of the polariton states: $LP_{i}$, $MP_{i}$ and $UP_{i}$, with $i = {nER, ER}$. The dashed ones correspond to bare light-hole ($X_\mathrm{lh}$) and heavy-hole ($X_\mathrm{hh}$) excitons as well as the cavity modes in the respective regions $C_\mathrm{nER}$ and $C_\mathrm{ER}$. The yellow vertical arrow denotes the position of the measured trap of the main text.  
        }
\label{Fig_SI_spatialDispersion}
\end{figure*}

\justifying
The polariton modes of the sample arise from to the strong coupling between the light-hole ($X_\mathrm{lh}$) and heavy-hole ($X_\mathrm{hh}$) excitons and cavity photon modes in the spatially extended non-etched ($C_\mathrm{nER}$) and etched ($C_\mathrm{ER}$) regions of the cavity~\cite{Kuznetsov2023}. Hence, three polariton modes are expected in each region. Figure~SI~\ref{Fig_SI_spatialDispersion} shows superimposed spatial dispersions of the polariton resonances in ER and nER across the sample. For each region the dispersions have been fitted using three coupled oscillator model to obtain the values of the bare resonances. The values corresponding to the position of the measured polariton trap are summarized in the Supplementary Table~\ref{polaritonDispersionFitParams}.

\begin{table}[htbp]
  \caption{Parameters determined from fits of  the spatial dispersion of the polaritons in extended etched (ER) and non-etched (nER) cavity regions.}
  \begin{tabular}{| c | c  | c |}
 \hline
 \hline
 Parameter   &  Value  &   Comment \\
 \hline
 $X_\mathrm{hh}$     &   $1532$~meV    &   Bare heavy-hole exciton energy  \\ 
 $X_\mathrm{lh}$     &   $1538$~meV    &   Bare light-hole exciton energy  \\ 
 $C_\mathrm{nER}$     &   $1519$~meV    & Bare cavity mode energy in non-etched region at $r = 4$~mm \\ 
 $C_\mathrm{ER}$     &   $1533$~meV    &   Bare cavity mode energy in etched region at $r = 4$~mm \\ 
 $\Omega_{\mathrm{X}_\mathrm{hh}}$     &   $3.0 \pm 0.12$~meV    &  Rabi-coupling energy for cavity-heavy-hole (for both regions)  \\ 
 $\Omega_{\mathrm{X}_\mathrm{lh}}$     &   $1.0 \pm 0.12$~meV    &  Rabi-coupling energy for cavity-light-hole (for both regions)  \\ 
\hline    
\hline
\end{tabular}
\label{polaritonDispersionFitParams}
\end{table}

\newpage
\section{Determination of maximum exciton energy modulation amplitude}
\label{appendix:reference}

\begin{figure*}[tbhp]
	\centering
	\includegraphics[width=0.6\textwidth, keepaspectratio=true]{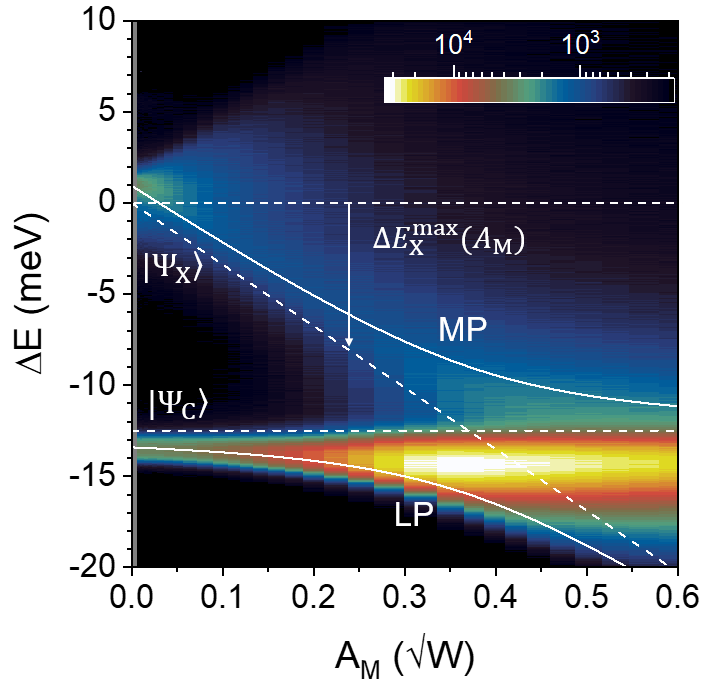}
	\caption{
        Dependence of the spectrum of a spatially extended non-etched cavity region (large negative detuning) on the acoustic modulation amplitude at low optical excitation power. The dashed diagonal line is the assumed maximum excursion of the bare exciton ($|\Psi_\mathrm{X}\rangle$) under the modulation. The dashed horizontal line is assumed energy of the bare photon mode ($|\Psi_\mathrm{C}\rangle$). The solid curves are fitted lower (LP) and middle (MP) polariton branches.
        }
\label{Fig_SI_ModulationAnticrosingFitting}
\end{figure*}

Figure~SI~\ref{Fig_SI_ModulationAnticrosingFitting} shows a dependence of the spectrum of a spatially extended non-etched cavity region vs. the acoustic modulation amplitude. The map was recorded at large negative detuning so that the lower polariton (LP) branch is photon-like, while the middle (MP) and the upper branches are exciton-like. Note that the upper branch located at about +5.5~meV is approximately 20 times weaker than the middle polariton at +1~meV. In the following we focus on the dominant polariton modes: LP \& MP.

In the strong coupling regime, only polariton modes can be observed in PL. Acoustic modulation offers a unique way to determine the dependence of exciton state $|\Psi_{X}\rangle$ energy ($E_{X}$) via the modulation of the polariton modes from the evolution of the trap spectrum as a function of the modulation amplitude ($A_\mathrm{M}$). In the first approximation, we can identify the downward slope of the maximum excursion of the bare exciton energy, $\Delta E_{X}^{max}(A_\mathrm{M})$, by simulating its anti-crossing with the bare photon ($|\Psi_\mathrm{C}\rangle$). In a simple case, the polariton modes are obtained from the simple two coupled oscillator model with a fixed coupling $\Omega_R$ between $|\Psi_\mathrm{X}\rangle$ and $|\Psi_\mathrm{C}\rangle$ as a function of the modulation amplitude ($A_\mathrm{M}$):
 
\begin{equation}
	H(A_\mathrm{M}) =\left(
		\begin{array}{cc}
		 \Delta_{0}+\Delta E_\mathrm{X}^\mathrm{max}(A_\mathrm{M}) & \Omega_\mathrm{R} \\
		 \Omega_\mathrm{R} & 0 \\
		\end{array}
		\right).
	\label{SI_EqHMP} 
\end{equation}

In the above, the energy is referenced to the unmodulated bare exciton one, which is set as zero. In the Hamiltonian, $\Delta_{0}$ is the static cavity-exciton detuning, $\Delta E_\mathrm{X}^\mathrm{max}(A_\mathrm{M})$ is the maximal shift of the exciton energy, which depends on the modulation amplitude.

The simulated polariton states are shown in the Figure~SI~\ref{Fig_SI_ModulationAnticrosingFitting} by white solid curves, which match the edges of energy-modulated LP and MP. The diagonal white dashed line indicates the fitted $\Delta E_\mathrm{X}^\mathrm{max}(A_\mathrm{M})$. The fitted coupling energy $\hbar \Omega_\mathrm{R} = 3.5$~meV matches closely the one determined in SI.~\ref{appendix:spatialDispersion}. The same procedure was used to obtain the crossing point in Fig.~1c of the main text.

\newpage
\section{Multi-mode condensation of confined polaritons}
\label{appendix:reference}

\begin{figure*}[tbhp]
	\centering
	\includegraphics[width=0.7\textwidth, keepaspectratio=true]{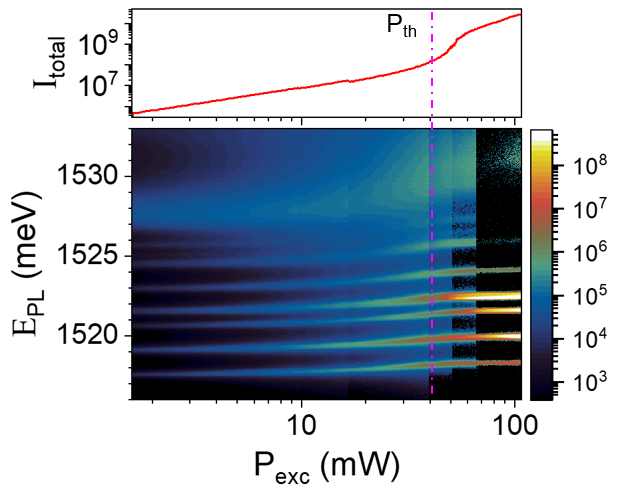}
	\caption{
        (bottom part) Evolution of the spatially integrated emission spectrum of a trap, studied in the main text, as a function of optical excitation power ($P_\mathrm{exc}$) of a non-resonant laser. (top part) Energy-integrated intensity of the trap as the function of $P_\mathrm{exc}$. The vertical dashed line denotes the multi-mode condensation threshold ($P_\mathrm{th}$).
        }
\label{Fig_SI_Condensation}
\end{figure*}

The $4 \times 4~\mu m^2$ trap discussed in the main text provides deep confinement of approximately -10~meV, given by the difference between the low polariton energies in the etched and non-etched regions as follows from Fig.~SI~\ref{Fig_SI_spatialDispersion}. Figure~SI~\ref{Fig_SI_Condensation} shows a dependence of the trap spectrum on the power ($P_\mathrm{exc}$) of a CW non-resonant laser. For very low $P_\mathrm{exc}$, the trap emission is spread uniformly among the confined levels. As the $P_\mathrm{exc}$ increases there is a continuous blueshift of the emission lines. When $P_\mathrm{exc} \approx P_\mathrm{th}$, the total intensity of the trap emission increases exponentially. When $P_\mathrm{exc} > P_\mathrm{th}$, the trap emission comes from several excited levels (the ground state has weak emission), which corresponds to the multi-mode polariton condensation regime~\cite{Eastham2008,Grosso2014}.

\newpage
\section{Acoustic control of confined condensates}
\label{appendix:reference}

\begin{figure*}[tbhp]
	\centering
	\includegraphics[width=1\textwidth, keepaspectratio=true]{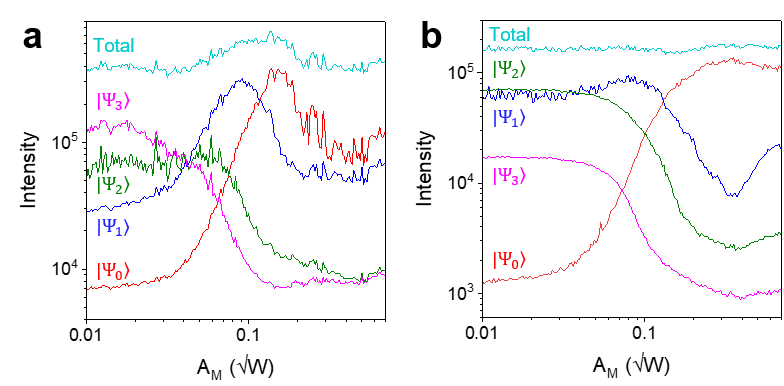}
	\caption{
        Dependence of the integrated intensities of different confined states and the total intensity of all of the levels as a function of the modulation amplitude ($A_\mathrm{M}$) for two optical excitation powers ($P_\mathrm{exc}$) relative to the threshold ($P_\mathrm{th}$, see Fig.~SI~\ref{Fig_SI_Condensation}): \bf{a} $P_\mathrm{exc} \approx P_\mathrm{th}$, \bf{b} $P_\mathrm{exc} \approx 2.3P_\mathrm{th}$.
        }
\label{Fig_SI_intensityVsAmplitude}
\end{figure*}

As discussed in the main text, large amplitude acoustic modulation has a profound effect on the spectral distribution of condensates emission intensity in the confined levels. Figure~SI~\ref{Fig_SI_intensityVsAmplitude} shows a dependence of the integrated intensities of different confined states as a function of the modulation amplitude for two powers ($P_\mathrm{exc}$) of a CW non-resonant laser relative to the BEC threshold ($P_\mathrm{th}$ determined for the modulation off condition, cf. Fig.~SI~\ref{Fig_SI_Condensation}): $\bf{a}$ $P_\mathrm{exc} \approx P_\mathrm{th}$ and $\bf{b}$ $P_\mathrm{exc} \approx 2.3P_\mathrm{th}$. These figures complement Fig.~2b of the main text for $P_\mathrm{exc} \approx 1.7P_\mathrm{th}$ clearly showing the qualitatively similar behavior of the condensates under the acoustic modulation, namely, progressive transfer of emission to the GS with increasing $A_\mathrm{M}$.

\newpage
\section{Linewidth dependence}
\label{appendix:reference}

\begin{figure*}[tbhp]
	\centering
	\includegraphics[width=0.6\textwidth, keepaspectratio=true]{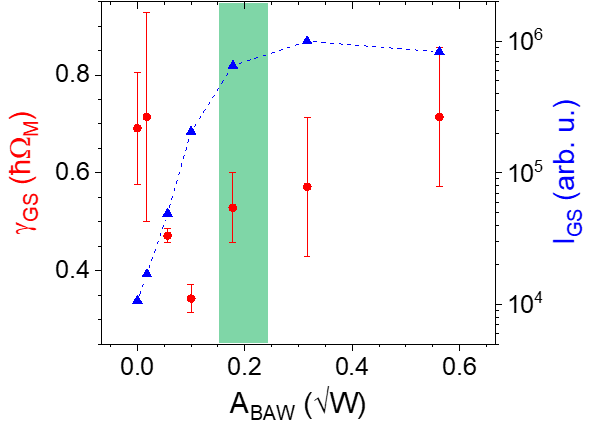}
	\caption{
        Dependence of the ground state linewidth (red circles), $\gamma_\mathrm{GS}$, expressed in the modulation quantum ($\hbar \Omega_\mathrm{M}$) and its integrated intensity, $I_\mathrm{GS}$, (blue triangles) on the modulation amplitude ($A_\mathrm{M}$) extracted from the Fig.~2c in the main text. The error bars give the standard deviation. The green rectangle indicates the amplitude range of the expected crossing of the bare (photonic) GS of the trap by the exciton (see Fig.~1c of the main text).    
        }
\label{Fig_SI_LinewidthVsAbaw}
\end{figure*}

As discussed in the main text, above the condensation threshold in the absence of the modulation, the ground state (GS) of the trap is weakly populated, hence its linewidth is relatively broad, cf. Fig.~2d of the main text. The red circles of the Fig.~SI~\ref{Fig_SI_LinewidthVsAbaw} show a dependence of the GS linewidth ($\gamma_\mathrm{GS}$), defined as the width of Lorentzians used to fit the spectra of Fig.~2d of the main text, on the modulation amplitude. The blue triangles show the dependence of the GS integrated intensity on $A_\mathrm{M}$. We see that for the small $A_\mathrm{M}$, the linewidth sharply decreases, which correlates with the exponential increase of the GS intensity. The linewidth starts to increase rapidly once the modulation amplitude exceeds the value corresponding to the expected crossing of the GS by the exciton (indicated by the green rectangle).

\newpage
\section{Auto-correlation setup}
\label{appendix:reference}

\begin{figure*}[tbhp]
	\centering
	\includegraphics[width=1\textwidth, keepaspectratio=true]{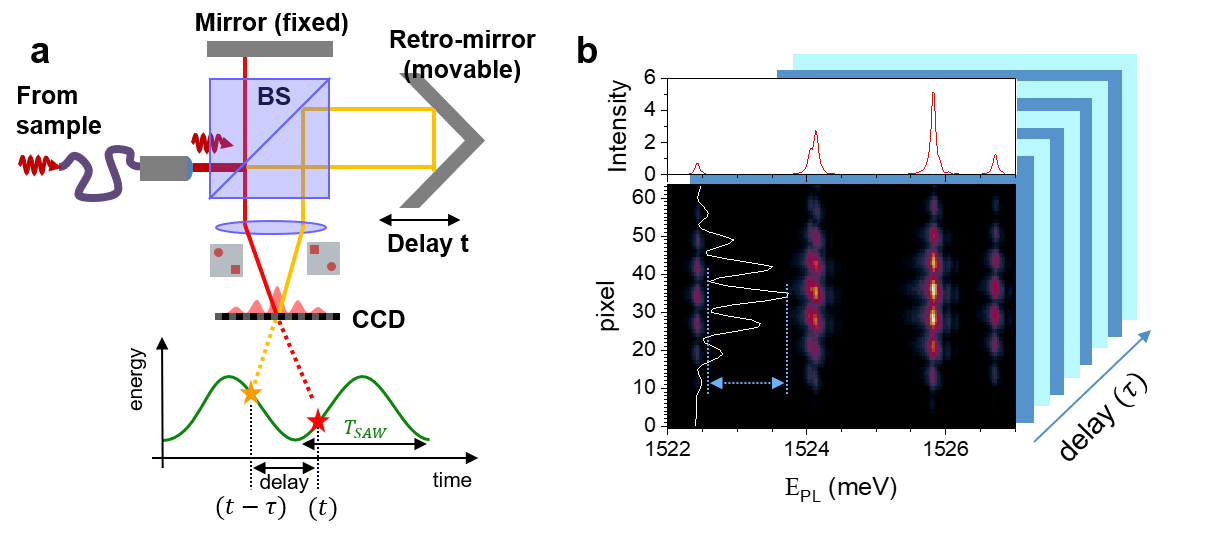}
	\caption{
        $\bf{a}$ Schematic representation of the interferometer-based setup for the measurement of the auto-correlation. In the figure, $BS$ is the 50/50 beam-splitter, $CCD$ is the a 2D array detector. 
        $\bf{b}$ A measured energy-resolved interferogram of a trap excited above the condensation threshold without the modulation and for zero-time delay ($\tau$). The white curve corresponds to the energy-integrated intensity of one of the states used to extract the maximum visibility, given by the dashed blue lines and an arrow. 
        }
\label{Fig_SI_CorrelationSetup}
\end{figure*}

Auto-correlation measurements shown in Fig.~3b of the main text were carried out using an interferometer schematically shown in Fig.~SI~\ref{Fig_SI_CorrelationSetup}a. The emission from the sample is first coupled into an optical fiber leading to the interferometer. At the fiber output, the light is collimated and the beam from the sample is 50/50 split and then directed into two arms of the interferometer. One arm has a fixed length. The other arm has a movable retro-reflector. The spatial shift of the latter translates into a time delay between the two beams. The latter are interfered on the 2D CCD detector.

Fig.~SI~\ref{Fig_SI_CorrelationSetup}b shows an exemplary spectrally resolved interferogram of the condensate at zero time delay and in the absence of modulation. We obtain the interference pattern by integrating over a selected energy range. From the pattern we extract the maximum visibility defined as $(I_\mathrm{max}-I_\mathrm{min})/(I_\mathrm{max}+I_\mathrm{min})$, which is the measure of the first order temporal coherence, $g^{(1)}$. The same method has recently been used to study the emergence of acceleration-induced beats of a polariton condensate modulated by sub-GHz surface acoustic waves~\cite{Kuznetsov2024}.

\newpage
\section{Auto-correlation for the excited state}
\label{appendix:reference}

\begin{figure*}[tbhp]
	\centering
	\includegraphics[width=0.8\textwidth, keepaspectratio=true]{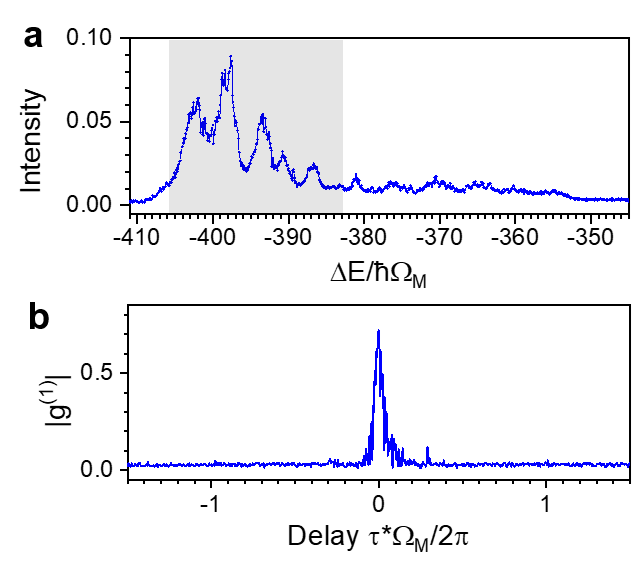}
	\caption{
        $\bf{a}$ Experimental spectrum of the excited state corresponding to the conditions in Fig.~3a in the main text. The energy scale is relative to the bare exciton energy and normalized to the modulation quantum $\hbar \Omega_\mathrm{M}$. $\bf{b}$ First order correlation function in units of the modulation frequency ($\Omega_\mathrm{M}/2\pi$) of the energy range defined by the gray background in $\bf{a}$.         
        }
\label{Fig_SI_CorrelationExcitedState}
\end{figure*}

As discussed in the main text (cf. Fig.~2a\&b), for large modulation amplitudes, excited states are significantly depleted, while the ground state contains most of the BEC population. In this case, the GS spectrum is a frequency comb of narrow lines separated by the modulation quantum (cf. main Fig.~3a). Figure~SI~\ref{Fig_SI_CorrelationExcitedState}a shows a spectrum of the first excited state recorded for the same conditions as the GS spectrum of Fig.~3a (main text). Due to the low polariton population, individual modulation sidebands can not be resolved, only the sidebands bunching (its origins is discussed in the main text). As expected, the $|g^{(1)}|$ of the excited state shows only a narrow single peak around zero delay, in contrast to the GS (cf. main Fig.~3b) showing correlations at integer multiples of the modulation period (cf. main Fig.~3b).

\newpage
\section{The model}\label{sec1}

In this section, we provide a more detailed account of the phenomenological model introduced in the main text, focusing on the key mechanisms that enable selective condensation in individual confined photonic levels. The model describes the time evolution of the polariton populations under strong periodic modulation of the exciton energy, and captures the interplay between coherent light-matter coupling, bosonic stimulation, and energy-selective injection. We identify three dominant mechanisms that drive the population dynamics within the polariton trap:

\begin{itemize} 
\item \textit{Resonant injection via strong light-matter coupling:} When the bare exciton energy is dynamically tuned into resonance with a confined photonic mode—within a detuning range of approximately $3$–$4 \ \mathrm{meV}$, comparable to the Rabi coupling energy -- the exciton-photon coupling enables an abrupt and efficient transfer of population into the corresponding polariton mode. This process is enabled by the strong strain-induced modulation, which periodically sweeps the exciton energy across multiple confined levels. 

\item \textit{Stimulated scattering:} Excitons relax into the confined polariton states through scattering processes. This relaxation is enhanced by bosonic stimulation, which promotes the accumulation of population in already-occupied modes, favoring the formation of macroscopically occupied condensate states. The scattering rate into each excited state depends on several factors, including the spatial overlap between the excitonic reservoir and the confined modes, as well as the excitonic Hopfield coefficient of the mode. i.e., the instantaneous energy detuning between the exciton level and each photonic mode.

\item \textit{Suppression of the scattering rate and enhancement of losses:} When the exciton energy is tuned below the confined photonic modes (corresponding to what we define as positive detuning), the polariton population in those modes becomes depleted. First, the scattering into polariton modes becomes inefficient, as relaxation processes are strongly suppressed under these conditions ~\cite{Wouters2007, Eastham2008, Grosso2014, Wouters2010}. Second, under strong strain modulation, the continuum of higher-energy exciton states can shift into resonance with the confined photonic modes. This overlap opens additional dissipation channels, significantly increasing the photon loss rate and further depleting the population in the affected modes.

\end{itemize}
\subsection{Resonant injection via strong light-matter coupling}\label{S1}

The microcavities studied in this work are specifically engineered to operate in the strong light-matter coupling regime. In particular, the Rabi splitting is estimated to be $\hbar \Omega_\mathrm{R} \sim 3\ \mathrm{meV}$ (see SI.~\ref{appendix:spatialDispersion}), indicating a pronounced hybridization between the excitonic and photonic components. To achieve lateral confinement of the photonic modes, the cavity spacer layer is etched such that localized regions possess a greater spacer thickness. This modification leads to spatial regions with a lower cavity photon energy, resulting in the formation of zero-dimensional polariton traps. A trap supports a discrete set of confined photonic modes, labeled $\psi_{\mathrm{C},j}$ ($j =  0, \dots, n-1$), where $n$ denotes the total number of confined levels supported by the trap.

We model the strong coupling between the exciton and the confined photonic modes using a simplified set of coupled equations that describe the time evolution of the exciton and photonic amplitudes:

\begin{eqnarray}
i\hbar \frac{d\psi_\mathrm{X}}{dt} &=&  \varepsilon_\mathrm{X} (t)  \psi_\mathrm{X} - \sum^{n}_{j=1} J_j \psi_{\mathrm{C},j} + \frac{i\hbar}{2}\left(P-\gamma_\mathrm{X} - R\alpha_\mathrm{X} \left|\psi_\mathrm{X}\right|^2\right) \psi_\mathrm{X}, \label{ESM1}\\
%
i\hbar \frac{d\psi_{\mathrm{C},j}}{dt} &=& \varepsilon_{\mathrm{C},j} \psi_{\mathrm{C},j} - J_j \psi_\mathrm{X} -\frac{i\hbar \gamma_{\mathrm{C},j}}{2} \psi_{\mathrm{C},j}\qquad \forall j.     
\label{ESM2}
\end{eqnarray}
Here, $\psi_\mathrm{X}$ represents the amplitude of the excitonic component, while $\varepsilon_\mathrm{X}$ and $\varepsilon_{\mathrm{C},j}$ are the bare energies of the exciton and the $j$-th confined photonic mode, respectively. The Rabi coupling energy between the exciton and each confined photonic mode $\psi_{\mathrm{C},j}$ is given by $J_j = \beta_j\hbar\Omega_\mathrm{R}/2$ ($\beta_j\leq1$), which determines the population exchange rate between the excitonic and photonic components for that specific mode.

We have also included the driven-dissipative terms that account for the effects of non-resonant pumping and particle losses. The term $P$ denotes the injection rate of excitons due to the pump, while $\gamma_\mathrm{X}$ and $\gamma_{\mathrm{C},j}$ represent the decay rates of the exciton and the $j$-th confined photonic mode, respectively. Additionally, a saturation term $R\alpha_\mathrm{X} \left|\psi_\mathrm{X}\right|^2$ is included, which effectively limits the exciton population due to phase-space filling effects. 

It is important to note that, at this stage, we do not include any direct stimulated scattering processes from the exciton reservoir into the photonic modes; the only mechanism for transferring population between excitons and photons is the coherent light-matter coupling term $J_j$.

For simplicity, this model considers a single effective excitonic state that interacts with each confined photonic mode through a mode-dependent coupling strength $J_j$. In reality, the exciton exhibits a nearly flat dispersion in the in-plane momentum space, with an energy that remains approximately constant across different momentum values. Each confined photonic mode primarily couples to the exciton state with the matching in-plane momentum. Under non-resonant excitation using a spatially localized laser with a Gaussian spot profile that is displaced with respect to the center of the trap, the population of excitons with different in-plane momenta is not uniform within the trap. This momentum-selective population is effectively captured in the model by assigning a different coupling constant $J_j$ for each mode. As we will show, this level of approximation is sufficient to reproduce the key features observed in the experiment.

The final and central element of our model is the dynamic modulation of the exciton energy in time, enabled by piezoelectric transducers. These transducers generate a coherent bulk acoustic wave that periodically strains semiconductor quantum wells of the microcavity, resulting in a time-periodic shift of the exciton energy. As a result, we consider a harmonic modulation of the form.
\begin{equation}
\varepsilon_\mathrm{X}(t) = \varepsilon_\mathrm{X} + A_\mathrm{M} \cos{\left(\Omega_\mathrm{M} t\right)},
\label{ESM3}
\end{equation}
where $A_\mathrm{M}$ is the energy modulation amplitude and $\Omega_\mathrm{M}$ is the acoustic driving frequency. In our case, the modulation amplitude can be made sufficiently large such that the exciton energy crosses all of the confined photonic levels within each oscillation cycle, satisfying the condition $A_\mathrm{M} > \left( \varepsilon_\mathrm{X} - \varepsilon_{\mathrm{C},0} \right)$. This strong periodic driving is the key mechanism that enables selective and time-dependent population transfer among the confined polariton states.

\begin{figure}[h]
\centering
\includegraphics[width=1\textwidth]{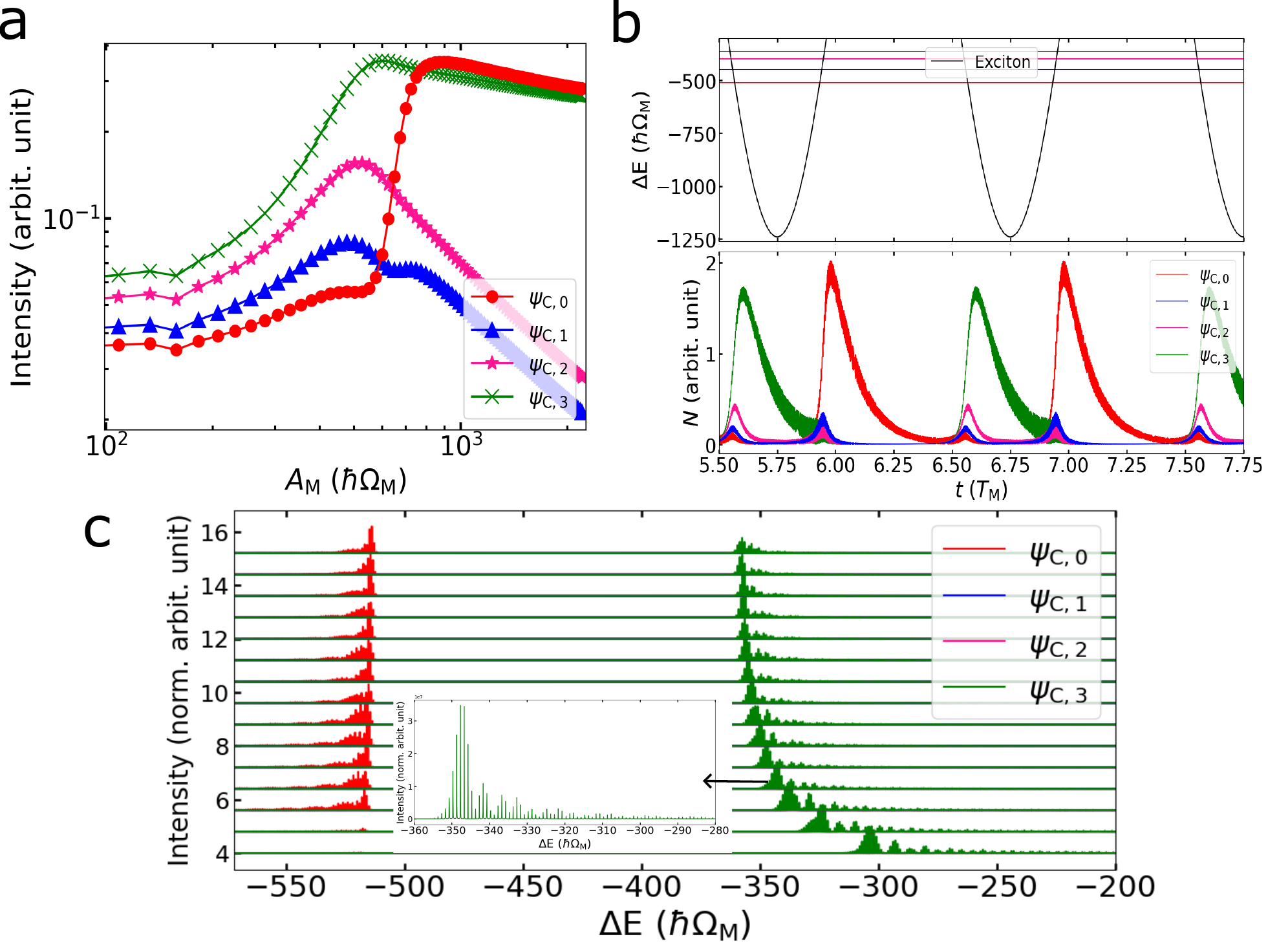}
\caption{\textbf{Resonant injection and spectral response under strong exciton modulation.}
\textbf{a} Integrated population of each confined photonic mode $\psi_{\mathrm{C},j}$ as a function of modulation amplitude $A_\mathrm{M}$, expressed in units of the phonon energy, calculated using the basic model from Eqs.~\ref{ESM1} and \ref{ESM2} with uniform Rabi coupling $J_j = J = 2\ \mathrm{meV}$.
\textbf{b} Top: Time evolution of the exciton energy relative to the confined modes (horizontal lines, color-coded as in panel \textbf{a}), with energy in units of $\hbar\Omega_\mathrm{M}$ and time in units of the modulation period $T_\mathrm{M}$. Zero energy corresponds to the unmodulated exciton level. Bottom: Time-dependent population $N_j = \left| \psi_{\mathrm{C},j} \right|^2$, showing pulsed emission driven by Landau-Zener-like population transfer during the exciton energy crossings.
\textbf{c} Emission spectra of the confined modes as a function of increasing $A_\mathrm{M}$, shown in cascade format. The modulation amplitude increases by $100\ \hbar \Omega_\mathrm{M}$ per spectrum from bottom to top. Each spectrum displays spectral sidebands with spacing $\omega_\mathrm{M}$, arising from phase and intensity modulation due to the periodic exciton energy drive. The resulting frequency comb structures deviate from standard Bessel-function envelopes due to the double-peaked temporal emission and anharmonic phase modulation. This deviation is illustrated in the inset, which corresponds to the fourth spectrum (from bottom to top) of mode $\psi_{\mathrm{C},3}$.
}

\label{Fig1SM}
\end{figure}

Figure~SI~\ref{Fig1SM}a shows the calculated population of each confined photonic mode ($\left| \psi_{\mathrm{C},j} \right|^2$) as a function of the modulation amplitude $A_\mathrm{M}$, obtained by numerically integrating Eqs.~\eqref{ESM1} and \eqref{ESM2}. In all simulations presented in this section, we assume a uniform light-matter coupling across modes, with $J_j = J = 2\ \mathrm{meV}$, corresponding to a Rabi splitting of $\hbar \Omega_\mathrm{R} = 4\ \mathrm{meV}$. In this example, the polariton trap supports four confined photonic modes, chosen to replicate conditions similar to those observed in the measured structures. The confined levels consist of the ground state (GS, $\psi_{\mathrm{C},0}$) at the lowest energy, followed by three excited states ($\psi_{\mathrm{C},1}$, $\psi_{\mathrm{C},2}$, and $\psi_{\mathrm{C},3}$), with $\psi_{\mathrm{C},3}$ being the highest in energy.

As the modulation amplitude $A_\mathrm{M}$ increases, the exciton energy is periodically swept over a broader spectral range. When the modulation is strong enough, it can span the full trap spectrum, crossing both below the GS and above the highest excited state. This mechanism leads to a strong injection of population into the confined modes as $A_\mathrm{M}$ increases. Specifically, from Fig.~SI~\ref{Fig1SM}a, the third excited state ($\psi_{\mathrm{C},3}$, green curve with $\times$-shaped symbols) is the first to gain significant population around $A_\mathrm{M} = 250\ \hbar\omega_\mathrm{M}$. Then, a sharp rise in the ground state population ($\psi_{C,0}$, red curve with circle symbols) occurs near $A_\mathrm{M} = 600\ \hbar\omega_\mathrm{M}$, when the modulation range extends below its energy.

At sufficiently large modulation amplitudes ($A_\mathrm{M} > 600\ \hbar\omega_\mathrm{M}$), the dominant emission arises from the highest and lowest energy modes in the trap: the third excited state and the ground state. This can be intuitively understood using a Landau-Zener-type picture. Initially, all population resides in the exciton reservoir, which undergoes periodic energy modulation and anticrosses with the confined photonic levels. Since the coupling strength ($\sim~2~\mathrm{meV}$) is much greater than the modulation frequency ($\hbar\omega_\mathrm{M} \ll J$), the system operates deep in the Landau-Zener (LZ) adiabatic regime, where the probability of population transfer during each avoided crossing approaches unity.

As the exciton energy sweeps downward, it first encounters the third excited state, transferring nearly all its population into this mode and leaving little for subsequent levels. Upon sweeping below the ground state, the exciton population is replenished by the continuous non-resonant pump. During the upward sweep, the exciton now encounters the GS first and efficiently transfers its population into it. This cyclic behavior explains why, at high modulation amplitudes, the population is concentrated primarily in the third excited state and the ground state.

This dynamic behavior is illustrated in Fig.~SI~\ref{Fig1SM}b. The top panel shows the time-dependent exciton energy sweeping across the photonic levels (indicated by horizontal colored lines). The bottom panel displays the corresponding time evolution of the populations of the photonic modes. The sudden population transfer due to the adiabatic LZ mechanism, followed by photon decay, results in a triangular-shaped pulse in time. Notably, the third excited state gains its population during the downward sweep of the exciton, while the ground state is populated during the upward sweep.

We also visualize the spectral consequences of the large modulation amplitudes in Fig.~SI~\ref{Fig1SM}c. The spectra are shown in a cascade format, where each trace is normalized to the maximum value across all four confined modes. The bottom-most spectrum corresponds to a modulation amplitude of $A_\mathrm{M} = 500\ \hbar\Omega_\mathrm{M}$, and each subsequent spectrum above corresponds to an increase of $100\ \hbar\Omega_\mathrm{M}$.

In the first three spectra (counting from the bottom), the third excited state dominates in intensity. Starting from the third trace, where the exciton energy begins to resonate with the ground state, both the third excited state and the ground state exhibit strong intensities. At higher modulation amplitudes, these two modes coexist with comparable and substantial emission.

A particularly striking feature in this figure is the appearance of multiple sidebands around both the third excited state and the ground state, with a separation corresponding to the phonon energy $\hbar\Omega_\mathrm{M}$. These sidebands form distinct spectral bundles whose spacing and structure vary with modulation amplitude. This behavior contrasts sharply with the typical sideband structure described by Bessel functions in standard frequency-modulated (FM) or amplitude-modulated (AM) signals. This is better visualized in the inset zooming in on one spectrum.

The origin of this deviation lies in the pulsed nature of the population dynamics of the photonic modes $\psi_{\mathrm{C},j}$, which results in temporal modulation patterns very different from those of conventional continuous waveforms. This pulsed behavior, along with its implications, is further discussed in the autocorrelation measurements presented in the main text.


\subsection{Stimulated scattering}\label{S2}

The model derived from Eqs.~\eqref{ESM1} and \eqref{ESM2} explains how lasing of the highest and lowest confined modes can be induced by increasing the modulation amplitude of the exciton energy. This demonstrates that the population dynamics within the trap can be significantly modified through the combined effects of strong light-matter coupling and large strain modulation in the microcavity. However, this simplified model overlooks a crucial feature observed in the experiments—namely, the appearance of lasing from the confined modes even in the absence of strain modulation.

To account for this, we propose an extended effective model governed by the following equations:

\begin{eqnarray}
i\hbar \frac{d\tilde{\psi}_\mathrm{X}}{dt} &=& \varepsilon_\mathrm{X}(t) \tilde{\psi}_\mathrm{X} - \sum^{n-1}_{j=0} J_j \tilde{\psi}_{\mathrm{C},j} + \frac{i\hbar \gamma_\mathrm{X}}{2}  \left( \tilde{P} - \alpha_\mathrm{X} \left| \tilde{\psi}_\mathrm{X}\right|^2 - \sum^{n-1}_{j=0} \alpha_j \left| \tilde{\psi}_{\mathrm{C},j}\right|^2\right) \tilde{\psi}_\mathrm{X}\,, \label{ESM4}\\
i\hbar \frac{d\tilde{\psi}_{\mathrm{C},j}}{dt} &=& \varepsilon_{\mathrm{C},j} \tilde{\psi}_{\mathrm{C},j} - J_j \tilde{\psi}_\mathrm{X} + \frac{i\hbar}{2}  \left( \gamma_\mathrm{X} \alpha_{j} \left| \tilde{\psi}_\mathrm{X}\right|^2 - \gamma_{\mathrm{C},j}\right) \tilde{\psi}_{\mathrm{C},j}\qquad\forall j\,,     \label{ESM5}
\end{eqnarray}
where we have introduced a rescaled set of variables: $\tilde{\psi}_\mathrm{X} = \psi_\mathrm{X}/\sqrt{\rho_0}$, $\tilde{\psi}_{\mathrm{C},j} = \psi_{\mathrm{C},j}/\sqrt{\rho_0}$, and $\tilde{P} = (P/\gamma_\mathrm{X}) - 1$, with $\rho_0 = \gamma_\mathrm{X}/R$.

We have included additional terms of the form $\alpha_j \left| \tilde{\psi}_{\mathrm{C},j} \right|^2$ to phenomenologically account for the various scattering mechanisms from the exciton reservoir into each confined photonic mode. These terms enable the description of lasing within the polariton trap even in the absence of strain modulation. Moreover, by tuning the relative values of the coefficients $\alpha_j$, we can control which confined states undergo lasing and which remain dark.

\begin{figure}[h]
\centering
\includegraphics[width=1\textwidth]{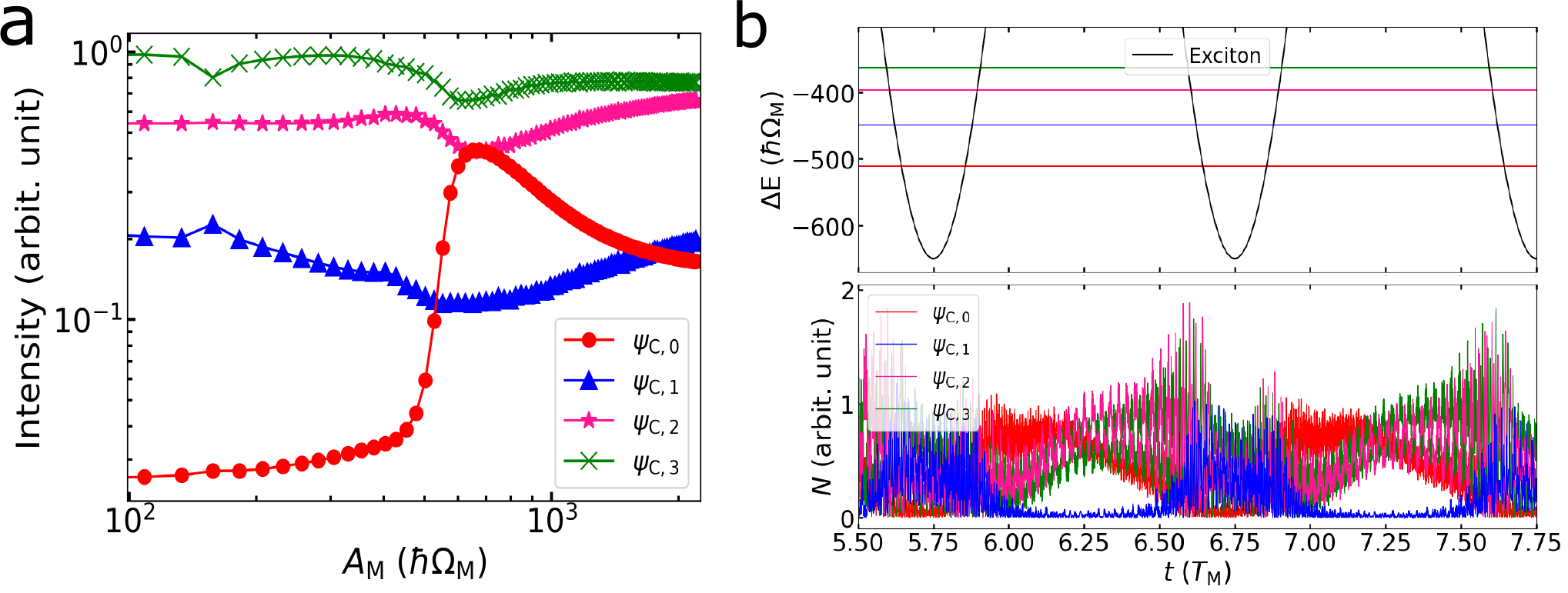}
\caption{
\textbf{a} Integrated population of each confined photonic mode $\psi_{\mathrm{C},j}$ as a function of the modulation amplitude $A_\mathrm{M}$, calculated using the extended model of Eqs.~\eqref{ESM4} and \eqref{ESM5}, which includes stimulated scattering terms $\alpha_j$. The scattering rates are chosen such that higher-energy modes (especially $\psi_{\mathrm{C},3}$ and $\psi_{\mathrm{C},2}$) dominate at low modulation amplitudes, replicating experimental conditions in which these modes lase preferentially in the absence of strain.
\textbf{b} Top: Time evolution of the exciton energy relative to the photonic levels (horizontal lines, color-coded as in panel \textbf{a}), showing the dynamic crossings during each oscillation cycle. Bottom: Time-resolved population dynamics $N_j = \left| \tilde{\psi}_{\mathrm{C},j} \right|^2$ of the four confined modes. While the pulsed structure of the emission is preserved, the model predicts simultaneous lasing from multiple modes and lacks the clean sequential depletion observed in experiments.
}
\label{Fig2SM}
\end{figure}

Figure~SI~\ref{Fig2SM}a shows the dependence of the mean population of each confined mode as a function of the strain modulation amplitude, $A_\mathrm{M}$, based on the extended model described in Eqs.~\eqref{ESM4} and \eqref{ESM5}. The parameters used in this simulation are the same as those in Fig.~SI~\ref{Fig1SM}, with the addition of the empirically tuned scattering coefficients: $\alpha_0 = 1.20$, $\alpha_1 = 1.80$, $\alpha_2 = 1.85$, and $\alpha_3 = 1.92$.

Under these conditions, and in the absence of exciton energy modulation ($A_\mathrm{M} = 0$), the population distribution favors the higher-energy confined modes: the third excited state ($j = 3$) is the most populated, followed by the second excited ($j = 2$), the first excited ($j = 1$), and finally the ground state ($j = 0$). This configuration is designed to match experimental conditions similar to those presented in the main text, where the highest-energy states exhibit the strongest emission. It also enhances the visibility of population redistribution effects driven by strain modulation.

It is worth noting that the stimulated scattering rates into the confined modes depend sensitively on various experimental factors, including the spatial overlap between the excitation spot and each orbital mode, as well as the energy detuning between the cavity modes and the exciton resonance (a topic we address in the following subsection). In this simplified model, these scattering rates are treated as effective parameters and adjusted empirically to reproduce the observed behavior of the measured traps.

As the modulation amplitude of the exciton energy increases, the most notable feature is the sharp rise in the ground state population, which occurs around the point where the modulated exciton energy approaches its bare value [$\left(\varepsilon_\mathrm{X} + A_\mathrm{M} - \varepsilon_{\mathrm{C},1}\right) \sim 2J$]. This sudden increase is a consequence of the Landau-Zener-like mechanism operating in the adiabatic regime, as discussed in the previous subsection.

The extended model thus captures both the multimode lasing in the confined modes via stimulated scattering and efficient population transfer to the ground state through strong light-matter coupling. However, it still falls short in reproducing the experimentally observed sequential filling and depletion of each excited state, which ultimately leads to dominant lasing in the ground state. Instead, the simulation predicts that, even at large modulation amplitudes, significant emission persists from both the third and second excited states alongside the ground state.
This behavior is illustrated in the temporal dynamics of these three modes, shown in Fig.~SI~\ref{Fig2SM}. While the pulsed nature of the emission is still present, the system exhibits a more complex interplay between the excited and ground states, deviating from the clean cascade behavior observed in the experiment.

\subsection{Suppression of the Scattering Rate and Enhancement of Losses}\label{S3}

To fully reproduce the experimental observations, a final ingredient must be incorporated into the phenomenological model: the energy-dependent behavior of both gain and loss associated with each confined photonic mode, relative to the exciton energy. In particular, the efficiency of stimulated scattering into each mode, as well as its radiative decay rate are known to depend strongly on the spectral detuning between the cavity mode and the exciton resonance~\cite{Wouters2007, Eastham2008, Grosso2014, Wouters2010}. Since the exciton energy evolves dynamically over time (see Eq.~\eqref{ESM3}), this detuning is time-dependent, and so are the scattering rates and losses of the confined modes.
We account for these effects through the following expressions:
\begin{eqnarray}
\alpha_{j}(t) &=& \frac{1}{2} \alpha_{j} \left\{1 - \tanh{\left[a \left( \varepsilon_{\mathrm{C},j} - \varepsilon_\mathrm{X}(t) \right) \right]} \right\}, \label{ESM6} \\
\gamma_{\mathrm{C},j}(t) &=& \gamma_{\mathrm{C},j} + \frac{1}{2} \gamma^{(\infty)} \left\{1 + \tanh{\left[b \left( \varepsilon_{\mathrm{C},j} - \varepsilon_\mathrm{X}(t) - \Delta\varepsilon_{ \mathrm{cont}} \right) \right]} \right\}. \label{ESM7}
\end{eqnarray}

Equation~\eqref{ESM6} models the time-dependent stimulated scattering rate from the exciton reservoir into each confined mode. Here, $\alpha_{j}$ is the baseline scattering strength (zero detuning), and $a$ is a smoothing parameter controlling the sharpness of the transition. The step-like form of $\alpha_{j}(t)$ ensures that when the exciton energy falls below the cavity mode energy, scattering into the mode is strongly suppressed -- capturing the fact that relaxation mechanisms become inefficient for positive detuning (exciton below the photonic mode).
This expression phenomenologically incorporates the dynamic nature of the gain in our system and allows the model to reproduce the selective filling of modes observed experimentally.

Equation~\eqref{ESM7} captures the sharp increase in the decay rate of the cavity modes when their energy approaches the continuum of unbound exciton states. This occurs when the modulation amplitude $A_\mathrm{M}$ is sufficiently large that the instantaneous exciton energy satisfies  $(\varepsilon_\mathrm{X}(t) + \Delta\varepsilon_{\mathrm{cont}}) < \varepsilon_{\mathrm{C},j}$, where $\Delta\varepsilon_{\mathrm{cont}}>0$ is the energy offset between the exciton ground state and the onset of the continuum band. In this regime, the overlap between the photonic modes and the continuum enhances the coupling to non-radiative decay channels, resulting in a significant increase in the effective loss rate.
This effect is phenomenologically modeled through a step-like increase in the dissipation term, governed by the hyperbolic tangent in Eq.~\eqref{ESM7}. The parameter $\gamma^{(\infty)}$ represents the additional loss induced by this coupling to the continuum, while $b$ is a smoothing parameter that controls the sharpness of the transition. The inclusion of this term allows the model to account for the experimentally observed suppression of lasing in cavity modes that become strongly positively detuned at large modulation amplitudes.

\begin{figure}[h]
\centering
\includegraphics[width=1\textwidth]{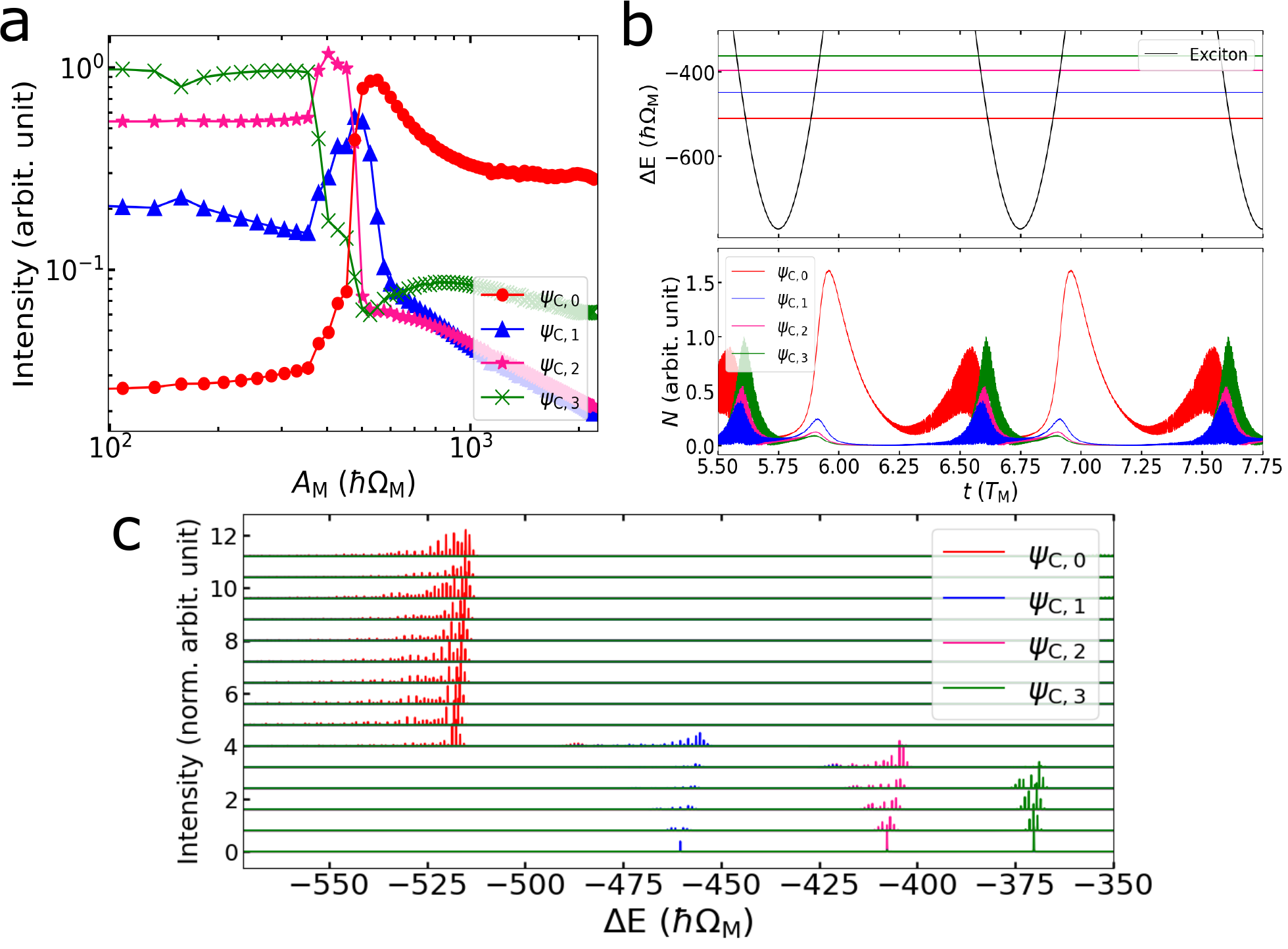}
\caption{
\textbf{a} Integrated population of each confined photonic mode $\psi_{\mathrm{C},j}$ as a function of the modulation amplitude $A_\mathrm{M}$, calculated using the full model incorporating time-dependent scattering rates $\alpha_j(t)$ and decay rates $\gamma_{\mathrm{C},j}(t)$. As $A_\mathrm{M}$ increases, a clear sequential depletion of higher-energy modes is observed, with population concentrating in the ground state $\psi_{\mathrm{C},0}$ at large modulation amplitudes. \textbf{b} Top: Time evolution of the exciton energy relative to the photonic modes (horizontal lines, color-coded as in panel \textbf{a}), showing the periodic crossings. Bottom: Corresponding time-dependent population $N_j = \left|\psi_{\mathrm{C},j}\right|^2$ for each mode. Unlike the previous version of the model, the ground state exhibits dominant and clean pulsed emission, while higher modes are strongly depleted due to detuning-dependent suppression of scattering and enhanced losses.
\textbf{c} Emission spectra of the confined modes as a function of increasing $A_\mathrm{M}$ (in cascade format). Each spectrum is vertically offset and corresponds to a $100\ \hbar\omega_\mathrm{M}$ increase in modulation amplitude from bottom to top. Zero energy corresponds to the unmodulated exciton energy. The spectral evolution confirms the sequential loss of higher-energy mode emission and the emergence of ground-state-dominated frequency combs, shaped by the double-peaked temporal dynamics and anharmonic light-matter hybridization.
}
\label{Fig3SM}
\end{figure}

Figure~SI~\ref{Fig3SM}a shows the integrated population of each confined photonic mode as a function of the strain modulation amplitude, $A_\mathrm{M}$, computed using the full model described in this section. The parameters used here are identical to those in Fig.~SI~\ref{Fig2SM}. As in the previous case, for low or vanishing modulation amplitudes, the higher-energy modes—specifically the second and third excited states—dominate the emission. This is a direct consequence of the chosen values for the stimulated scattering rates $\alpha_j$, which favor population of these modes in the absence of modulation.

As $A_\mathrm{M}$ increases, the exciton energy begins to dynamically sweep across the various confined modes. The combined effect of the three key mechanisms—stimulated scattering, radiative losses, and Landau-Zener-like adiabatic transfer leads to a sequential population and depletion of each mode as the exciton energy approaches and then moves past their respective resonances. This dynamic redistribution of the population is clearly seen in Fig.~SI~\ref{Fig3SM}a.

At high modulation amplitudes, when the minimum exciton energy $\min(\varepsilon_\mathrm{X}(t))$ drops below the energy of the ground mode ($\varepsilon_{\mathrm{C},0}$), the ground state emerges as the dominant lasing mode. The underlying time dynamics, shown in Fig.~SI~\ref{Fig3SM}b, reveal that population is efficiently transferred into the excited modes as the exciton energy sweeps downward. However, these modes are rapidly depleted as the exciton crosses below them. The ground state follows a similar trend during the down-sweep but displays a key distinction during the upward passage of the exciton energy: a significant fraction of the exciton population is transferred into the ground mode via strong light-matter coupling, reinforcing its dominance.

Moreover, among all confined modes, the ground state is the least affected by the energy-dependent suppression of gain and enhancement of losses. It spends the shortest amount of time in the regime where both its effective scattering rate is reduced and its radiative decay is amplified due to the reduction of the cavity quality factor. This gives the ground mode a competitive advantage in capturing and retaining exciton population, consolidating its role as the dominant lasing mode at high modulation amplitudes. Ultimately, this results in a strong, ground-state-dominated emission featuring a characteristic double-peaked pulse structure in time, corresponding to the exciton energy sweeping through the cavity resonance in both directions.

To complement the analysis, Fig.~SI~\ref{Fig3SM}c displays the spectral evolution of the cavity emission as a function of the modulation amplitude, $A_\mathrm{M}$, presented in a cascading format. The bottom spectrum corresponds to $A_\mathrm{M} = 0$, with each subsequent trace representing an increment of $100\ \hbar\Omega_\mathrm{M}$, and all spectra are normalized to their respective maximum intensities. These simulations clearly reveal the sequential population transfer from higher- to lower-energy modes as $A_\mathrm{M}$ increases. Moreover, the emission from the trapped modes exhibits frequency combs, which arise from the combined phase and intensity modulation induced by the periodic exciton energy drive at frequency $\Omega_\mathrm{M}/2\pi$. Similar to the observations in Fig.~SI~\ref{Fig1SM}c, the sideband distribution deviates significantly from the ideal Bessel function envelope. This deviation stems from the double-peaked temporal structure of the emission and the anharmonic phase modulation introduced by the time-dependent light-matter hybridization of the confined modes.

As a final note, it is relevant to highlight the choice of parameters and contrast them with those presented in the main text and in Subsection~\ref{sec1}D of the SI. In particular, throughout this section, the Rabi coupling is assumed to be uniform across all cavity modes ($J_j = J$), in contrast to the main text, where $J_j$ varies for each confined mode to account for the imbalance in the population of excitons with different in-plane momenta. Such imbalance can arise, for example, from a spatial displacement between the exciton reservoir and the center of the trap. This simplification ($J_j = J$) shows that the key experimental features can be qualitatively captured without fine-tuning the individual coupling constants. The more detailed comparison with experimental results presented in the main text employs mode-dependent coupling values $J_j$ to more accurately reproduce the observed dynamics and relative mode intensities.

\newpage
\subsection{Simulation parameters}
\label{appendix:reference}

Parameters of the simulation used to produce results shown in Fig.~2e \& Fig.~3c-e are shown in the Supplementary Table~\ref{simulationParams}.

\begin{table}[htbp]
  \caption{Parameters used in the simulation. Modulation frequency is $\Omega_\mathrm{M}/2\pi$. Condensation threshold power is $P_\mathrm{th}$.}
  \begin{tabular}{| c | c | c | c |}

 \hline
~~ Parameter   ~~&~~  Value  ~~&~~  Unit  ~~&~~ Comment ~~\\

 \hline
 $P_\mathrm{exc}$   &  5   &  $P_\mathrm{th}$  &   Power of non-resonant optical excitation\\
 \hline
 $A_\mathrm{M}$   &  variable   &  $\hbar \Omega_\mathrm{M}$  &   Peak amplitude of the sinusoidal modulation of the bare exciton energy\\

 \hline
 \hline
 $E_\mathrm{X}$   &  0   &  $\hbar \Omega_\mathrm{M}$  &   Bare exciton energy without modulation \\
 \hline
 $E_\mathrm{b}$   &  -135   &  $\hbar \Omega_\mathrm{M}$  &   Barrier energy with respect to the bare exciton\\
 \hline
 $E_{4}$   &  -359   &  $\hbar \Omega_\mathrm{M}$  &  $4^{th}$ confined state energy with respect to the bare exciton  \\
 \hline
 $E_{3}$   &  -411   &  $\hbar \Omega_\mathrm{M}$  &  $3^{rd}$ confined state energy with respect to the bare exciton  \\
 \hline
 $E_{2}$   &  -446   &  $\hbar \Omega_\mathrm{M}$  &  $2^{nd}$ confined state energy with respect to the bare exciton  \\
 \hline
 $E_{1}$   &  -498   &  $\hbar \Omega_\mathrm{M}$  &  $1^{st}$ confined state energy with respect to the bare exciton  \\  
 \hline
 $E_{0}$   &  -560   &  $\hbar \Omega_\mathrm{M}$  &  Ground state (GS) energy with respect to the bare exciton  \\ 

 \hline
 \hline
 $J_\mathrm{b}$   &  3   &  $\hbar \Omega_\mathrm{M}$  &  Barrier Rabi splitting energy  \\ 
 \hline
 $J_{4}$   &  12   &  $\hbar \Omega_\mathrm{M}$  &  $4^{th}$ confined state Rabi splitting energy  \\
 \hline
 $J_{3}$   &  69   &  $\hbar \Omega_\mathrm{M}$  &  $3^{rd}$ confined state Rabi splitting energy  \\
 \hline
 $J_{2}$   &  50   &  $\hbar \Omega_\mathrm{M}$  &  $2^{nd}$ confined state Rabi splitting energy  \\
 \hline
 $J_{1}$   &  19   &  $\hbar \Omega_\mathrm{M}$  &  $1^{st}$ confined state Rabi splitting energy  \\  
 \hline
 $J_{0}$   &  19   &  $\hbar \Omega_\mathrm{M}$  &  GS Rabi splitting energy  \\  

 \hline
 \hline
 $\alpha_\mathrm{X}$   &  2.4   &    &  Filling rate constant of the exciton from the optical pump \\
 \hline
 $\alpha_\mathrm{b}$   &  0   &    &  Filling rate constant of the barrier from the exciton\\
 \hline
 $\alpha_{4}$   &  1.6   &   &  Filling rate constant of the $4^{th}$ confined state from the exciton\\
 \hline
 $\alpha_{3}$   &  1.9   &    &  Filling rate constant of the $3^{rd}$ confined state from the exciton\\
 \hline
 $\alpha_{2}$   &  1.9   &    &  Filling rate constant of the $2^{nd}$ confined state from the exciton\\
 \hline
 $\alpha_{1}$   &  1.7   &   &  Filling rate constant of the $1^{st}$ confined state from the exciton\\
 \hline
 $\alpha_{0}$   &  1.6   &    &  Filling rate constant of the GS confined state from the exciton\\

 \hline
 \hline
 $\gamma_\mathrm{X}$   &  1.2   & $\hbar \Omega_\mathrm{M}$   &  Decay rate of the exciton \\
 \hline
 $\gamma_\mathrm{b}$   &  5   &  $\hbar \Omega_\mathrm{M}$  &  Decay rate of the barrier \\
 \hline
 $\gamma_{4}$   &  1.5   &  $\hbar \Omega_\mathrm{M}$ &  Decay rate of the $4^{th}$ confined state \\
 \hline
 $\gamma_{3}$   &  1.5   &  $\hbar \Omega_\mathrm{M}$  &  Decay rate of the $3^{rd}$ confined state \\
 \hline
 $\gamma_{2}$   &  1.5   &  $\hbar \Omega_\mathrm{M}$  &  Decay rate of the $2^{nd}$ confined state\\
 \hline
 $\gamma_{1}$   &  1.5   & $\hbar \Omega_\mathrm{M}$  &  Decay rate of the $1^{st}$ confined state \\
 \hline
 $\gamma_{0}$   &  1.5   &  $\hbar \Omega_\mathrm{M}$  &  Decay rate of the GS \\
 
\hline    

\end{tabular}
\label{simulationParams}
\end{table}

\newpage
\section{Simulation of the modulation}
\label{appendix:reference}

\begin{figure*}[tbhp]
	\centering
	\includegraphics[width=0.6\textwidth, keepaspectratio=true]{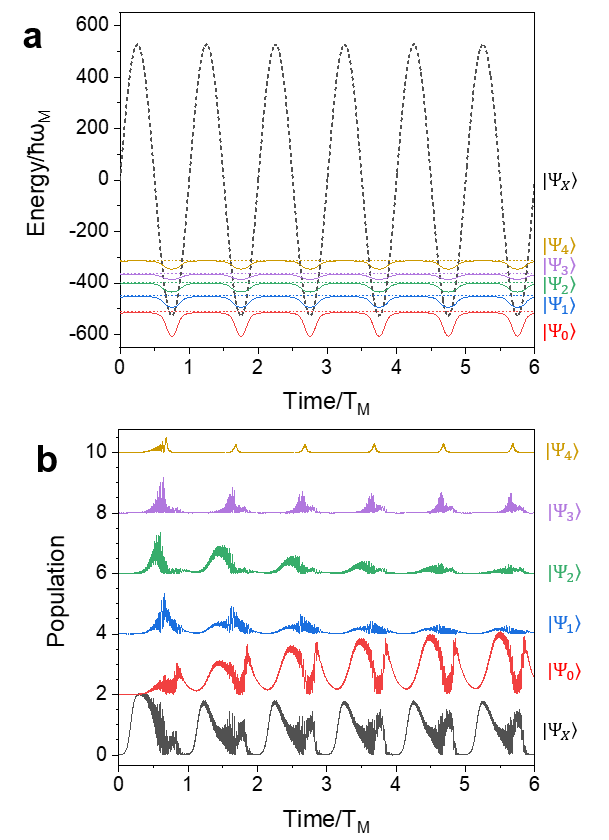}
	\caption{
        {\bf d} Simulated time-evolution of the energy of the bare exciton ($|\Psi_\mathrm{X}\rangle$) and confined photon states (dashed horizontal lines, $|\Psi_{i}\rangle$, with $i = 0$ corresponding to the ground state) for several modulation periods ($T_\mathrm{M}$). The zero-time corresponds to the turn-on of the modulation. The solid curves show the energies of the confined polariton states.
        {\bf b} The corresponding evolution of photonic level populations for the modulation conditions in the panel $\bf{a}$.        
        }
\label{Fig_SI_CalcTimeEvolution}
\end{figure*}

Figure SI~\ref{Fig_SI_CalcTimeEvolution}a shows the time-evolution of the instantaneous energies of the coupled exciton-photon system, taking into account only the Hamiltonian part of Eqs. \eqref{ESM4} and \eqref{ESM5}, over six modulation periods ($T_\mathrm{M}$) after the turn-on of the acoustic modulation. The energy levels configuration, excitation and the modulation amplitude approximately correspond to the conditions of Fig.~1c of the main text for $A_\mathrm{M} \approx 0.25$. Only the lower four polariton modes are shown. The black dashed line correspond to the modulation of the bare exciton energy. Figure SI~\ref{Fig_SI_CalcTimeEvolution}b shows the population of the bare photonic and excitonic modes for the full dissipative evolution. 

We consider the first $T_\mathrm{M}$ after the modulation turn-on: $0 \le t \le T_\mathrm{M}$. When the exciton energy is well above the confined photon states, the exciton holds most of the population and the levels are empty. As the detuning reduces, the higher levels become progressively more populated, while the exciton is de-populated. The maximum of the level population is achieved very close to the crossing point. The details of the population transfer are discussed in the main text. As the exciton moves below the bare photon ground state ($|\Psi_{0}\rangle$), the latter is considerably depopulated. We note that within the first few cycles, the peak value of ground state population increases and then stabilizes. This is not a numerical artifact, rather a consequence of the stimulated scattering mechanism whose efficiency depends on the final state population.

\newpage
\section{Simulated spectrum under modulation}
\label{appendix:reference}

\begin{figure*}[tbhp]
	\centering
	\includegraphics[width=1\textwidth, keepaspectratio=true]{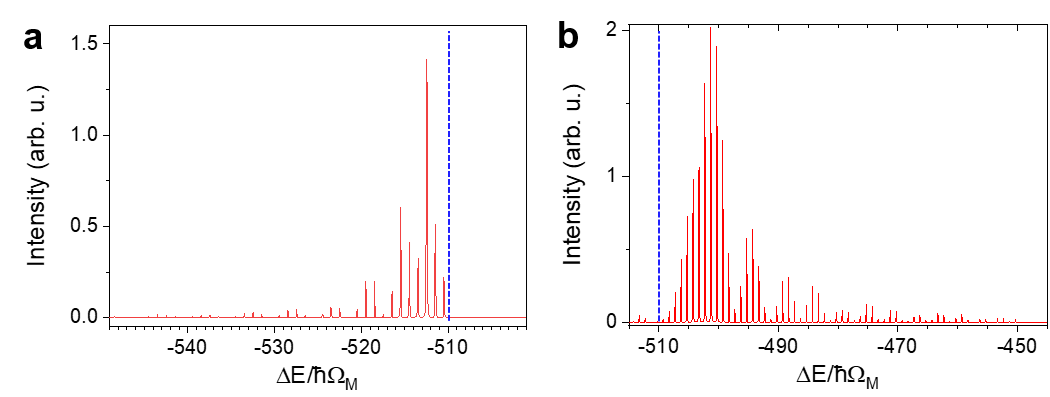}
	\caption{
        Polariton ground state spectra calculated based on the model for the two acoustic modulation amplitudes: $\bf{a}$ $A_\mathrm{M} = 797 * \hbar \Omega_\mathrm{M}$ and $\bf{b}$ $A_\mathrm{M} = 551 * \hbar \Omega_\mathrm{M}$. The modulation amplitude corresponds to the maximum peak shift of the bare exciton with respect to its unmodulated energy. The energy scale in both panels is referenced to the unmodulated exciton energy. The vertical dashed blue lines designate the position of the bare unmodulated bare photonic mode of the trap.    
        }
\label{Fig_SI_CalcSpectra}
\end{figure*}

The model shows, in different parameters regimes, the spectral features observed for the condensate under acoustic modulation. Figure~SI~\ref{Fig_SI_CalcSpectra} shows two ground state (GS) spectra calculated for modulation amplitudes: $\bf{a}$ $A_\mathrm{M} = 797 * \hbar \Omega_\mathrm{M}$ and $\bf{b}$ $A_\mathrm{M} = 551 * \hbar \Omega_\mathrm{M}$. Both spectra show well-resolved modulation sidebands and sidebands bunching (discussed in the main text). In this case, the only difference between Fig.~SI~\ref{Fig_SI_CalcSpectra}a and Fig.~SI~\ref{Fig_SI_CalcSpectra}b is the modulation amplitude. We observe that when the energy modulation amplitude considerably exceeds the separation between the bare exciton and GS ($\Delta E_\mathrm{GS}$), panel $\bf{a}$, the spectrum is asymmetric towards the higher energies. The opposite is generally true when $A_\mathrm{M} \approx \Delta E_\mathrm{GS}$, panel $\bf{b}$. Also, the model predicts that the domination of the stimulated scattering over the Rabi-transfer (e.g., small Rabi coupling rates) leads to the spectral asymmetry towards the higher energies. This can be understood as the consequence of the assumption of the model that the bosonic stimulation turns off, when the exciton is driven below a bare confined level. Finally, we note that the asymmetry of the intensity of the spectrum towards lower or higher energies depends sensitively on the simulation parameters.

\vskip 1 cm
\bibliography{Mendeley_Bib}

